\documentclass[11pt]{article}
\usepackage{amssymb,amsmath,amsfonts}
\usepackage{graphicx}
\usepackage{graphics}
\usepackage{eepic,epsfig}
\usepackage{dcolumn}

\@addtoreset{equation}{section}

\makeatother

\begin{document}

\title{Vacuum currents in curved tubes}
\author{A. A. Saharian \\
\textit{Institute of Physics, Yerevan State University, }\\
\textit{1 Alex Manoogian Street, 0025 Yerevan, Armenia }}
\maketitle

\begin{abstract}
We investigate the combined effects of spatial curvature and topology on the
properties of the vacuum state for a charged scalar field localized on
rotationally symmetric 2D curved tubes. For a general spatial geometry and
for quasiperiodicity condition with a general phase, the representation of
the Hadamard function is provided where the topological contribution is
explicitly extracted. As an important local characteristic of the vacuum
state the expectation value of the current density is studied. The vacuum
current is a periodic function of the magnetic flux enclosed by the tube
with the period of flux quantum. The general formula is specified for
constant radius and conical tubes. As another application, we consider the
Hadamard function and the vacuum current density for a scalar field on the
Beltrami pseudosphere. Several representations are provided for the
corresponding expectation value. For small values of the proper radius of
the tube, compared with the curvature radius, the effect of spatial
curvature on the vacuum current is weak and the leading term in the
corresponding expansion coincides with the current density on a constant
radius tube. The effect of curvature is essential for proper radii of the
tube larger than the radius of spatial curvature. In this limit the fall-off
of the current density, as a function of the proper radius, follows a
power-law for both massless and massive fields. This behavior is in clear
contrast to the one for a constant radius tube with exponential decay for
massive fields. We also compare the vacuum currents on the Beltrami
pseudosphere and on locally de Sitter and anti-de Sitter 2D tubes.
\end{abstract}

\bigskip


\bigskip

\section{Introduction}

The canonical quantization of fields is based on the expansion of the field
operator in terms of a complete set of mode functions being the solutions of
the classical field equation. The coefficients of the expansion determine
the annihilation and creation operators that are used for the construction
of the Fock space of states started from the vacuum state. The latter is
defined as the state of quantum field nullified by the action of the
annihilation operator. The mode functions are sensitive to both local and
global characteristics of the background space-time and the same is the case
for the properties of the vacuum and particle states. Among the interesting
areas of research in quantum field theory is the dependence of those
properties on the geometry and topology of the background spacetime. The
corresponding effects play an important role in gravity, cosmology, in
physical models with extra compact dimensions, in condensed matter physics
and in finite temperature field theory (see, for example, \cite%
{Birr82,Grib94,Park09}). As an example of the combined influence of the
spatial curvature and topology on the properties of quantum vacuum, in the
present paper we consider the generation of vacuum currents for a scalar
field localized on 2D curved tubes. These currents have common roots with
persistent currents in mesoscopic rings, widely studied in the literature
for different physical systems (see, e.g., \cite{Dunn99,Imry08,Fomi18,Saha24}
and references therein). The experimental detection of persistent currents
in normal metal rings has been reported in \cite{Bles09,Bluh09}.

The investigation of field theoretical effects in (2+1)-dimensional
spacetimes has attracted a great deal of attention. The interest is
motivated by several reasons. In addition to be simplified toy models of
(3+1)-dimensional physics, the change in the number of spatial dimension
leads to new types of phenomena. They include new mechanisms for symmetry
breaking, parity violation, and fractionalization of quantum numbers. From
the fundamental point of view, an important aspect in 2D gauge theories is
the mechanism that generates a topological mass for gauge bosons without
breaking the gauge symmetry \cite{Dese82}. 2D models also appear as
effective field theories describing the physics of compact subspace in
Kaluza-Klein and braneworld type theories with extra dimensions. The recent
advances in synthesis methods and techniques for planar condensed matter
systems (2D materials) have increased significantly the interest to 2D
physics. The long-wavelength degrees of freedom in a number of those systems
is described by 2D relativistic field theory. Well known examples are the
graphene family materials with the low-energy excitations of the electronic
subsystem described by 2D Dirac equation involving the Fermi velocity
instead of light velocity (see, e.g., \cite{Gusy07,Cast09}).

In the physics of 2D materials the topological issues play an important role
both in the space of states and in the coordinate space. In particular,
topological phases of matter and topological defects are currently the
subject of active research. Here we will investigate the influence of
spatial topology and curvature on the ground state properties of the 2D
scalar field. As the representative of those properties the current density
is taken. The simplest example of nontrivial topology is a cylindrical tube
with constant radius. The vacuum currents in that geometry for scalar and
Dirac fields have been investigated in \cite{Bell15sc,Bell10} as special
cases of more general geometries with an arbitrary number of spatial
dimensions and with toroidally compactified subspace. Both cases of infinite
and finite length tubes were considered. The corresponding finite
temperature charge and current densities have been discussed in \cite%
{Beze13T,Bell14T}. The vacuum currents induced by magnetic fluxes in conical
geometries, describing idealized cosmic strings, are considered in \cite%
{Srir01,Site09,Beze10cs,Site18}. More realistic geometry of a finite
thickness cosmic string with a cylindrically symmetric core is discussed in
\cite{Beze15}. The compactification of cosmic strings along the axial
direction may lead to the component of the current density along that
direction \cite{Beze13Comp,Brag15Comp}. The ground state charge and current
densities for a fermionic field on a two-dimensional circular ring were
studied in \cite{Bell16Ring}.

The geometric curvature of 2D materials provides an additional mechanism to
control their physical characteristics and opens new perspectives for
applications. Among the interesting features is the generation of effective
gauge fields \cite{Guin10,Levy10}. The investigation of the curvature
induced effects in those systems is also important from the point of view of
fundamental physics giving insights on the influence of gravitational field
on quantum matter. Motivated by these technological and fundamental
perspectives, the effects of curvature in 2D physics are in the focus of
active research (see, for example, \cite{Vozm10}-\cite{Wei23}, covering
various aspects of the topic). Related to the topic of the present paper,
the vacuum expectation value of the current density for scalar and fermionic
fields in locally de Sitter (dS) and anti-de Sitter (AdS) spacetimes with
toroidal subspaces is investigated in \cite{Bell13dS,Beze15AdS} (for a
review see \cite{Saha24}).

The paper is organized as follows. In the next section we describe the
background geometry and present some special cases. The scalar field modes
and the Hadamard function for a general rotationally symmetric tube are
given in Section \ref{sec:Modes}. By using the Hadamard function, the vacuum
expectation value of the current density is investigated in Section \ref%
{sec:Current}. The vacuum currents for some special cases of the tube
geometry are presented in Section \ref{sec:Special}. As an application of
general formulas, in Section \ref{sec:Beltrami} we investigate the
expectation value of the current density on the Beltrami pseudosphere. The
asymptotic analysis and numerical results are presented. The main results of
the are summarized in Section \ref{sec:Conc}. In Appendix \ref{sec:dStubes}
the expressions for the current densities in (2+1)-dimensional locally dS
and AdS spacetimes with a spatial dimension compactified to a circle are
presented.

\section{Geometry of the problem and special cases}

\label{sec:Geom}

\subsection{Background geometry and induced metric}

\begin{figure}[tbph]
\begin{center}
\epsfig{figure=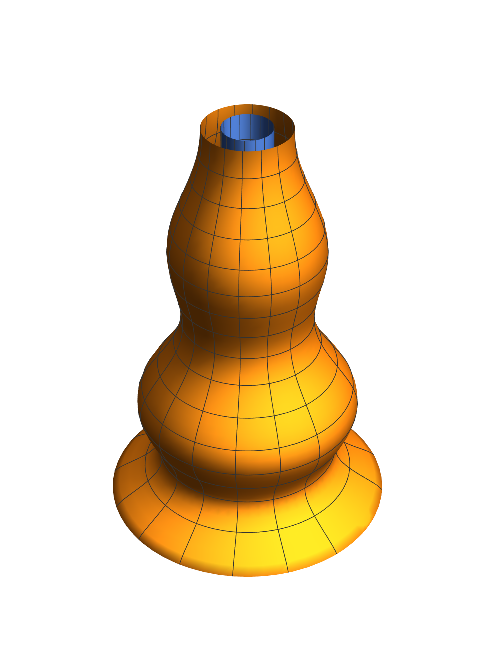,width=6cm,height=6cm}
\end{center}
\caption{Curved tube threaded by a magnetic flux.}
\label{fig1}
\end{figure}

We consider a cylindrical surface in a 3-dimensional Euclidean space with
coordinates $X^{i}=(X,Y,Z)$, $i=1,2,3$. The equation of the surface in the
parametric form is expressed as%
\begin{equation}
X=f(u)\cos \phi ,\;Y=f(u)\sin \phi ,\;Z=z(u),  \label{Par1}
\end{equation}%
where $\phi $ is the angular coordinate, $0\leq \phi \leq 2\pi $. The radius
of the surface for a fixed $Z$ is given by the function $f(u)=\sqrt{%
X^{2}+Y^{2}}$ (see Fig. \ref{fig1}). In the corresponding (3+1)-dimensional
Minkowski spacetime the line element has the form $ds_{4}^{2}=\eta _{\mu \nu
}dX^{\mu }dX^{\nu }$, where $\eta _{\mu \nu }$ is the Minkowskian metric
tensor. We will denote the spacetime coordinates on the cylindrical surface
by $x^{i}=(x^{0}=t=X^{0},x^{1}=u,x^{2}=\phi )$. For the line element on the
surface one has $ds^{2}=g_{ik}dx^{i}dx^{k}$, where the components of the
induced metric $g_{ik}=\eta _{\mu \nu }\partial _{i}X^{\mu }\partial
_{k}X^{\nu }$ are given by
\begin{equation}
g_{00}=1,\;g_{11}=-\left[ f^{\prime 2}(u)+z^{\prime 2}(u)\right]
,\;g_{22}=-f^{2}(u),  \label{g00}
\end{equation}%
and the off-diagonal components vanish. Here the prime stands for the
derivative of the function with respect to the argument.

The metric tensor can be further simplified introducing a new coordinate $w$
in accordance with
\begin{equation}
w=w(u)=\int_{u_{0}}^{u}du\sqrt{f^{\prime 2}(u)+z^{\prime 2}(u)}.  \label{w}
\end{equation}%
In terms of this coordinate the line element on the cylinder is written as
\begin{equation}
ds^{2}=dt^{2}-dw^{2}-p^{2}(w)d\phi ^{2},  \label{ds33}
\end{equation}%
where the function $p(w)$ is defined by the relation $p(w)=f(u(w))$, with $%
u=u(w)$ determined from (\ref{w}). By taking into account that $p^{\prime
}(w)=1/\sqrt{1+z^{\prime 2}(u)/f^{\prime 2}(u)}$, for the function $%
h(w)=z(u(w))$ one gets
\begin{equation}
h(w)=\pm \int_{w_{0}}^{w}dw\sqrt{1-p^{\prime 2}(w)},  \label{hu}
\end{equation}%
with $w_{0}$ being a constant. In terms of $w$ the equation of the surface
in the parametric form is rewritten as
\begin{equation}
X=p(w)\cos \phi ,\;Y=p(w)\sin \phi ,\;Z=h(w),  \label{Par2}
\end{equation}%
where the function $h(w)$ is defined by (\ref{hu}). In coordinates $%
(t,w,\phi )$, the nonzero components of the Ricci tensor and the Ricci
scalar for a (2+1)-dimensional spacetime determined by (\ref{ds33}) have the
form
\begin{equation}
R_{1}^{1}=R_{2}^{2}=-\frac{p^{\prime \prime }(w)}{p(w)},\;R=-2\frac{%
p^{\prime \prime }(w)}{p(w)}.  \label{RBirr}
\end{equation}%
The radius of the tube for a fixed $w$ is given by $p(w)$ and for the length
of compact dimension we have $2\pi p(w)$.

\subsection{Special cases}

Let us consider some special cases of the general geometry described above.
The simplest one corresponds to a cylinder with constant radius $L$. In this
case one has
\begin{equation}
p(w)=L  \label{pcyl}
\end{equation}%
and $-\infty <w<+\infty $. The second example is described by a linear
function%
\begin{equation}
p(w)=\alpha w,\;\alpha =\mathrm{const}.  \label{pc}
\end{equation}%
Assuming $\alpha <1$, for the function $h(w)$ from (\ref{hu}) one finds%
\begin{equation}
h(w)=\sqrt{1-\alpha ^{2}}w.  \label{hc}
\end{equation}%
where we have taken $w_{0}=0$. The line element takes the form%
\begin{equation}
ds^{2}=dt^{2}-dw^{2}-\alpha ^{2}w^{2}d\phi ^{2},  \label{ds2c}
\end{equation}%
with $0\leq w<\infty $. For $\alpha =1$ this is the line element on a plane
in polar coordinates. For $\alpha <1$ it describes a cone with an opening
angle $2\pi \alpha $. It is obtained from a plane cutting the angle $2\pi
\left( 1-\alpha \right) $ and gluing the edges of the remaining angle. For a
given coordinate $Z=h(w)$ in the embedding space the radius of the circle is
given by (\ref{pc}). The tip of the cone is located at $(X,Y,Z)=(0,0,0)$.
Note that for $w\neq 0$ one has $R_{ik}=0$ and both the spacetime and space
are flat. The graphitic cones are among the possible condensed matter
realizations of 2D conical geometry. The corresponding values of the
parameter $\alpha $ are determined by the structure of the graphene
hexagonal lattice and are given by $\alpha =1-N_{c}/6$ with $N_{c}=1,2,3,4,5$%
. Graphitic cones with all those values of the parameter $\alpha $ have been
observed in experiments (see, e.g., \cite{Kris97,Naes09}).

As the next example we consider a constant curvature space with
\begin{equation}
\frac{p^{\prime \prime }(w)}{p(w)}=\pm \frac{1}{a^{2}}=\mathrm{const}.
\label{pcc1}
\end{equation}%
For the corresponding Gaussian curvature one has $K=R/2=\mp 1/a^{2}$. In the
case of a constant positive curvature space (lower sign in (\ref{pcc1})) the
solution can be taken in the form%
\begin{equation}
p(w)=L\sin \left( w/a\right) ,\;0\leq w/a\leq \pi .  \label{pccp}
\end{equation}%
Introducing the angular coordinate $\theta =w/a$, $0\leq \theta \leq \pi $,
the line element is written as%
\begin{equation}
ds^{2}=dt^{2}-a^{2}\left( d\theta ^{2}+\sin ^{2}\theta d\phi ^{2}\right) .
\label{S2}
\end{equation}%
This corresponds to a sphere with radius $a$. The condensed matter
realizations of this geometry include fullerenes and topological insulators
with spherical surfaces \cite{Vozm10,Gonz93,Lee09,Taka13}. In the
long-wavelength approximation the dynamics of the corresponding electronic
subsystem is described by the 2D Dirac equation with a sphere as a
background space.

For a negative curvature space (the upper sign) one gets the equation%
\begin{equation}
p^{\prime \prime }(w)=p(w)/a^{2}.  \label{pcng}
\end{equation}%
Three separate subcases are realized. For the first one we take the solution
of (\ref{pcng}) given by%
\begin{equation}
p(w)=Le^{w/a},\;-\infty <w<+\infty ,  \label{pexp}
\end{equation}%
where $L$ is a constant with dimension of length. For the line element on
the tube one gets%
\begin{equation}
ds^{2}=dt^{2}-dw^{2}-L^{2}e^{2w/a}d\phi ^{2}.  \label{ds2BPS}
\end{equation}%
The spatial geometry described by this element corresponds to the Beltrami
pseudosphere. The coordinates of the embedding space are given by (\ref{Par2}%
). For the corresponding part of the surface we have $p^{\prime }(w)\leq 1$.
This shows that though in (\ref{pexp}) one has $-\infty <w<+\infty $, only
the part of the space corresponding to
\begin{equation}
w\leq w_{\mathrm{b}}\equiv -a\ln (L/a),  \label{wlimB}
\end{equation}%
can be embedded in a 3-dimensional Euclidean space. For the function $h(w)$
in the embedding space (with the upper sign in (\ref{hu})) one gets%
\begin{equation}
h(w)=a\left[ \sqrt{1-e^{2\left( w-w_{\mathrm{b}}\right) /a}}-\mathrm{arctanh}%
\left( \sqrt{1-e^{2\left( w-w_{\mathrm{b}}\right) /a}}\right) \right] .
\label{hwB}
\end{equation}%
On the left panel of Fig. \ref{fig1} we have presented the spatial geometry
corresponding to (\ref{ds2BPS}) as a surface (\ref{Par2}) in an Euclidean
space with coordinates $(X,Y,Z)$.
\begin{figure}[tbph]
\begin{center}
\begin{tabular}{ccc}
\epsfig{figure=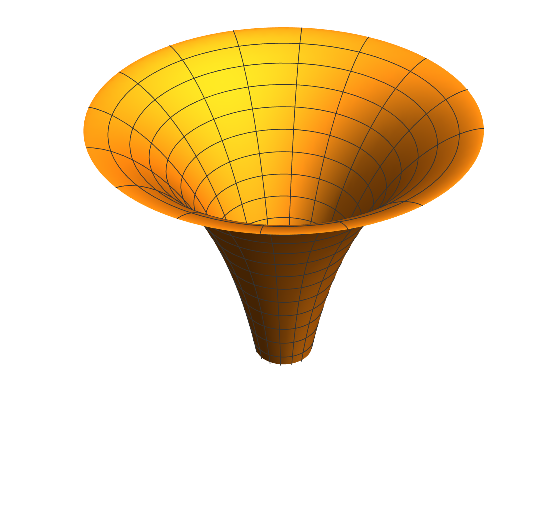,width=5cm,height=5cm} & %
\epsfig{figure=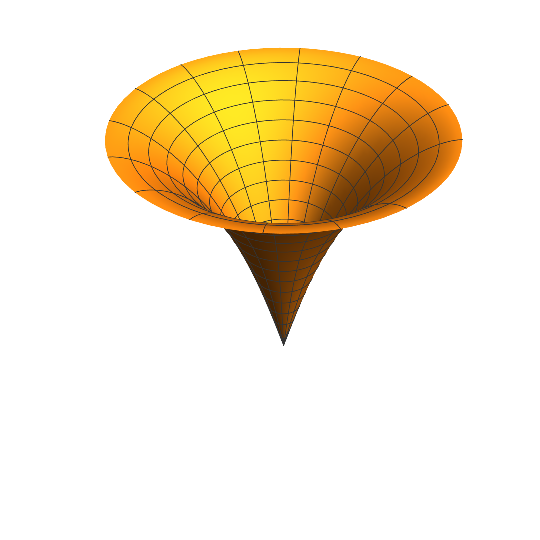,width=5.5cm,height=5.5cm} & %
\epsfig{figure=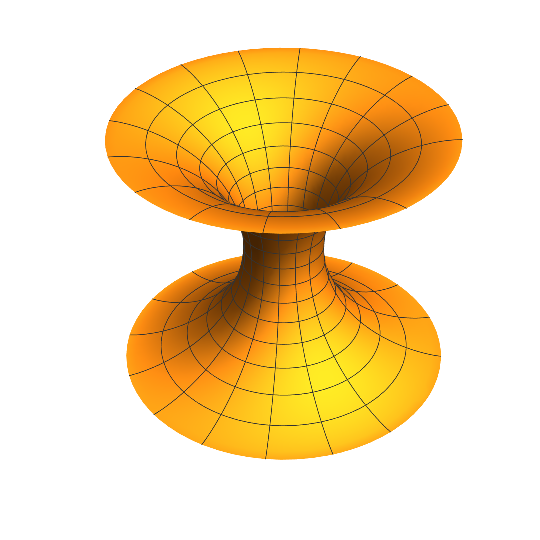,width=5cm,height=5cm}%
\end{tabular}%
\end{center}
\caption{The Beltrami (left), elliptic (middle) and hyperbolic (right)
pseudospheres embedded in a 3-dimensional Euclidean space.}
\label{fig2}
\end{figure}

For the second subcase of the negative curvature space we take
\begin{equation}
p(w)=L\sinh (w/a),\;0\leq w<\infty ,  \label{pES}
\end{equation}%
with the line element%
\begin{equation}
ds^{2}=dt^{2}-dw^{2}-L^{2}\sinh ^{2}(w/a)d\phi ^{2}.  \label{ds2ES}
\end{equation}%
The spatial geometry corresponds to the elliptic pseudosphere. For the part
embedded in a 3-dimensional Euclidean space one finds%
\begin{equation}
0\leq w\leq w_{\mathrm{be}}\equiv a\,\mathrm{arccosh}(a/L),  \label{wlimEP}
\end{equation}%
with $L/a\leq 1$ and the function $h(w)$ in (\ref{Par2}) is given by
\begin{equation}
h(w)=a\int_{0}^{w/a}dx\sqrt{1-(L/a)^{2}\cosh ^{2}x}.  \label{hwES}
\end{equation}%
The integral in the right-hand side is expressed in terms of the elliptic
functions. The middle panel in Fig. \ref{fig2} presents the part of the
elliptic pseudosphere embedded in an Euclidean space.

The third subcase for the negative curvature space corresponds to the
function
\begin{equation}
p(w)=L\cosh (w/a),\;-\infty <w<\infty ,  \label{pHS}
\end{equation}%
and the line element is given by
\begin{equation}
ds^{2}=dt^{2}-dw^{2}-L^{2}\cosh ^{2}(w/a)d\phi ^{2}.  \label{ds2HS}
\end{equation}%
This correspond to the hyperbolic pseudosphere. For the piece of the
geometry embedded in a Euclidean space we have
\begin{equation}
-w_{\mathrm{bh}}\leq w\leq w_{\mathrm{bh}}\equiv a\,\mathrm{arcsinh}(a/L).
\label{wlimHP}
\end{equation}%
The function $h(w)$ is determined from
\begin{equation}
h(w)=a\int_{0}^{w/a}dx\sqrt{1-(L/a)^{2}\sinh ^{2}x}.  \label{hwHS}
\end{equation}%
Similar to the previous case this integral is expressed in terms of the
elliptic functions. On the right panel of Fig. \ref{fig2} we display the
part of the hyperbolic pseudosphere embedded in Euclidean space in
accordance with (\ref{Par2}).

The condensed matter realizations of negative constant curvature surfaces by
2D materials have been discussed by several authors. In particular, those
structures have been used as simplified models for 2D black holes and
wormholes (see, for example, \cite%
{Gonz10,Iori12,Cvet12,Iori14,Taio16,Capo18,Rojj19,Kand20,Iori21,Alen21,Gall21}
and references therein). One can get new kinds of structures combining
different geometries discussed above. For example, cylindrical and planar
geometries have been combined to construct simplified models for 2D
wormholes. The combination of spherical and cylindrical geometries is
realized in capped graphene nanotubes. The vacuum expectation values of the
field squared and of the energy-momentum tensor for a massive scalar field
in the latter geometry have been studied in \cite{Beze16}.

\section{Scalar field modes and the Hadamard function}

\label{sec:Modes}

Having specified the geometry we turn to the field. We consider a complex
scalar field $\varphi (x)$ localized on a cylindrical surface. The
corresponding Lagrangian density reads
\begin{equation}
\mathcal{L}=g^{kl}\left( D_{k}\varphi \right) ^{\dagger }\left( D_{l}\varphi
\right) -\left( \xi R+m^{2}\right) \varphi ^{\dagger }\varphi
,\;D_{k}=\nabla _{k}+ieA_{k},  \label{Lag}
\end{equation}%
where the dagger stands for the hermitian conjugate, $A_{k}$ is the vector
potential for a classical gauge field, $\xi $ is the curvature coupling
parameter and $\nabla _{k}$ is the covariant derivative operator
corresponding to the metric tensor $g^{ik}$ determined by (\ref{ds33}). In
the special cases of minimally and conformally coupled scalar fields one has
$\xi =0$ and $\xi =1/8$, respectively. The complex scalar field describes
charged excitations in Bose-Einstein condensates, phononic degrees of
freedom and surface plasmons (see \cite{Leiz17,Muss19,Jung21} for some
recent references). The necessity for an investigation into the dynamics of
fields in two-dimensional space also arises in the context of holographic
models. Those models establish duality betwee two theories living in
spacetimes with different spatial dimensions. A well-studied example in the
literature is the duality between the string theory or supergravity in $%
(D+1) $-dimensional AdS spacetime and $D$-dimensional conformal field theory
(CFT) living on its boundary (AdS/CFT correspondence, for reviews see \cite%
{Ahar00,Nast15}). This type of correspondence enables the investigation of
non-perturbative effects in a given theory by examining the effects in the
weak coupling region of the dual theory. In particular, the AdS/CFT
correspondence for $D=2$ has been employed for the investigation of strong
coupling problems in condensed matter physics, including quantum phase
transitions, topological insulators, and holographic superconductors (see,
e.g., \cite{Zaan15}).

The field equation obtained from the variational principle with the action $%
S=\int d^{3}x\,\sqrt{g}\mathcal{L}$, $g=\mathrm{det}\,(g_{k l})$, is given
by
\begin{equation}
\left( g^{ik}D_{i}D_{k}+m^{2}+\xi R\right) \varphi (x)=0.  \label{Feq}
\end{equation}
We consider a background with nontrivial topology and the periodicity
condition along the direction of the angular coordinate $\phi $ needs to be
specified. We will impose a quasiperiodicity condition
\begin{equation}
\varphi (t,w,\phi +2\pi )=e^{i\tilde{\alpha}_{p}}\varphi (t,w,\phi ),
\label{Per0}
\end{equation}%
with a constant phase $\tilde{\alpha}_{p}$. For the gauge field a simple
configuration will be considered with $A_{k}=(0,0,A_{2}=\mathrm{const})$ in
the spacetime coordinates $(t,w,\phi )$. This corresponds to a magnetic flux
$\Phi =-2\pi A_{2}$ in the embedding space, threading the tube. On the tube,
the field tensor for the gauge field is zero and the effect on the
properties of 2D quantum scalar field is of the Aharonov-Bohm type. The
gauge field under consideration is excluded from the field equation by the
gauge transformation $\varphi ^{\prime }(x)=e^{ie\varkappa (x)}\varphi (x)$,$%
\;A_{k}^{\prime }=A_{k}-\partial _{k}\varkappa (x)$ with the function $%
\varkappa (x)=A_{2}\phi $. In the new gauge $A_{k}^{\prime }=0$ and the
scalar field obeys the periodicity condition%
\begin{equation}
\varphi ^{\prime }(t,w,\phi +2\pi )=e^{i\alpha _{p}}\varphi ^{\prime
}(t,w,\phi ),\;\alpha _{p}=\tilde{\alpha}_{p}+2\pi eA_{2}.  \label{Per}
\end{equation}%
The phase shift in this condition is presented as $2\pi eA_{2}=-2\pi \Phi
/\Phi _{0}$, where $\Phi _{0}=2\pi /e$ is the flux quantum. The discussion
below will be presented in terms of the field $\varphi ^{\prime }(x)$
omitting the prime.

The canonical quantization procedure for scalar fields is standard (see,
e.g., \cite{Birr82,Park09} for general background geometries and in general
number of spatial dimensions and \cite{Niko03} for the case of charged
fields). The canonically conjugate momenta for the fields $\varphi (x)$ and $%
\varphi ^{\dagger }(x)$ are given by $\pi _{k}(x)=\partial \mathcal{L}%
/\partial (D^{k}\varphi )=\left( D_{k}\varphi \right) ^{\dagger }$ and $\pi
_{k}^{\dagger }(x)=\partial \mathcal{L}/\partial (D^{k}\varphi )^{\dagger
}=D_{k}\varphi $. Taking a spacelike surface $\Sigma $ with unit normal $%
n^{k}$ and introducing the operators $\pi (x)=n^{k}\pi _{k}(x)$ and $\pi
^{\dagger }(x)=n^{k}\pi _{k}^{\dagger }(x)$, the canonical commutation
relations read (see also \cite{Niko03})
\begin{eqnarray}
\lbrack \varphi ^{\dagger }(x),\varphi (x^{\prime })]_{\Sigma } &=&[\pi
^{\dagger }(x),\pi (x^{\prime })]_{\Sigma }=0,  \notag \\
\lbrack \varphi ^{\dagger }(x),\pi (x^{\prime })]_{\Sigma } &=&[\varphi
(x),\pi ^{\dagger }(x^{\prime })]_{\Sigma }=0,  \notag \\
\lbrack \varphi (x),\pi (x^{\prime })]_{\Sigma } &=&[\varphi ^{\dagger
}(x),\pi ^{\dagger }(x^{\prime })]_{\Sigma }=i\delta (x,x^{\prime })/\sqrt{%
\gamma },\;x,x^{\prime }\in \Sigma ,  \label{ComRel}
\end{eqnarray}%
where $\gamma $ is the determinant of the induced metric on the surface $%
\Sigma $. In the problem under consideration the space-time is static (see
\cite{Dowk78} for general quantization procedure in static spacetimes) and
for $\Sigma $ we can take $t=\mathrm{const}$ with $n^{k}=(1,0,0)$ and $%
\gamma =p^{2}(w)$. To construct the Fock space of particle states the field
operator is expanded in terms of a complete set of orthonormalized mode
functions $\{\varphi _{\sigma }^{(+)}(x),\varphi _{\sigma }^{(-)}(x)\}$
being the solutions of the classical field equation and specified by the
collective set of quantum numbers $\sigma $. The expansion reads
\begin{equation}
\varphi (x)=\sum_{\sigma }\left[ a_{\sigma }\varphi _{\sigma
}^{(+)}(x)+b_{\sigma }^{\dagger }\varphi _{\sigma }^{(-)}(x)\right] ,
\label{phiexp}
\end{equation}%
where the symbol $\sum_{\sigma }$ means the summation over discrete quantum
numbers and the integration over continuous ones in the set $\sigma $. The
coefficients $a_{\sigma }$ and $b_{\sigma }^{\dagger }$ in (\ref{phiexp})
are the annihilation and creation operators for particles with quantum
numbers $\sigma $. From (\ref{ComRel}) one gets the standard commutation
relations $[a_{\sigma },a_{\sigma ^{\prime }}^{\dagger }]=[b_{\sigma
},b_{\sigma ^{\prime }}^{\dagger }]=\delta _{\sigma \sigma ^{\prime }}$ and
the commutators for other pairs of operators vanish. Here, $\delta _{\sigma
\sigma ^{\prime }}$ is understood as the Kronecker delta for discrete
quantum numbers and as the Dirac delta function for continuous ones. The
vacuum state of a quantum field is defined by the relations $a_{\sigma
}|0\rangle =b_{\sigma }|0\rangle =0$ for all values of quantum numbers $%
\sigma $. Then, the $a$- and $b$-particle states are constructed in the
standard way, acting on the vacuum state by the creation operators $%
a_{\sigma }^{\dagger }$ and $b_{\sigma }^{\dagger }$.

The construction described above shows that the notions of vacuum and
particle, in general, depend on the choice of the orthonormal set of mode
functions. Taking the second set of modes $\{\bar{\varphi}_{\chi }^{(+)}(x),%
\bar{\varphi}_{\chi }^{(-)}(x)\}$, with the new set of quantum numbers $\chi
$, we can write the expansion (Bogolubov transformations) $\bar{\varphi}%
_{\chi }^{(s)}=\sum_{\sigma }(\alpha _{\chi \sigma }^{(s)}\varphi _{\sigma
}^{(+)}+\beta _{\chi \sigma }^{(s)}\varphi _{\sigma }^{(-)})$ for $s=+,-$.
It can be shown that the Fock spaces based on the modes $\varphi _{\sigma
}^{(\pm )}(x)$ and $\bar{\varphi}_{\chi }^{(\pm )}(x)$ are different if the
Bogolubov coefficient $\beta _{\chi \sigma }^{(s)}\neq 0$ (see \cite%
{Birr82,Grib94,Park09} for a general discussion). In particular, the vacuum
state $|\bar{0}\rangle $ for the modes $\bar{\varphi}_{\chi }^{(\pm )}(x)$
will contain particles of the modes $\varphi _{\sigma }^{(\pm )}(x)$. Well
known examples of inequivalent vacuum states in flat spacetime in the
absence of external fields are the Minkowski and Fulling-Rindler vacua. They
are the vacuum states for inertial and uniformly accelerating observers,
respectively. In static geometries with the time coordinate $x^{0}=t$ the
spacetime possesses a global Killing vector field $\partial _{t}$. This
allows to define the positive and negative energy mode functions in
accordance with $\partial _{t}\varphi _{\sigma }^{(+)}=-i\omega \varphi
_{\sigma }^{(+)}$ and $\partial _{t}\varphi _{\sigma }^{(-)}=i\omega \varphi
_{\sigma }^{(-)}$, where $\omega $ is the single-particle energy. The most
natural vacuum state in those geometries is the state based on the canonical
quantization in terms of those modes. The particle detectors (in particular
of the Unruh-deWitt type) at rest in the corresponding reference frame will
remain in their ground state (no particles are detected).

The properties of the vacuum state $|0\rangle $ for the field $\varphi (x)$
under consideration are contained in two-point functions describing the
correlations of the vacuum fluctuations in different spacetime points. For
the evaluation of two-point functions we need to have the complete set of
positive and negative energy scalar modes $\{\varphi _{\sigma
}^{(+)}(x),\varphi _{\sigma }^{(-)}(x)\}$ obeying the field equation (\ref%
{Feq}) with $A_{k}=0$ and the condition (\ref{Per}). For the geometry
described by (\ref{ds33}) one has two Killing vectors $\partial _{t}$ and $%
\partial _{\phi }$ and the mode functions possessing these symmetries are
presented in the form%
\begin{equation}
\varphi _{\sigma }^{(\pm )}(x)=e^{ik_{n}\phi \mp i\omega t}z_{\sigma
}(w),\;k_{n}=n+\frac{\alpha _{p}}{2\pi },\;n=0,\pm 1,\pm 2,\ldots ,
\label{phisol}
\end{equation}%
where the eigenvalues of the momentum $k_{n}$ along the azimuthal direction
are determined by the condition (\ref{Per}). For an observer with fixed
coordinates $w$ and $\phi $ the basis (\ref{phisol}) presents a natural set
for the quantization of the field and for the choice of the vacuum state (as
natural as the plane-wave modes for the quantization of fields in Minkowski
spacetime). In particular, the particle detectors (for example, of the
Unruh-DeWitt type, see \cite{Birr82}) at rest in the reference frame
determined by the coordinates $(t,w,\phi )$ will not be excited if the field
is prepared in the vacuum state corresponding to the modes (\ref{phisol}).
In addition, being specified by conserved quantum numbers $\omega $ and $%
k_{n}$, those modes define a vacuum state having the same degree of symmetry
as the background metric tensor (examples of this type of vacuum states are
the Minkowski vacuum in flat spacetime and the Bunch-Davies vacuum in de
Sitter spacetime). The same class of the modes has been used previously in a
number of special cases, such as cylindrical and conical tubes.

Substituting the mode functions (\ref{phisol}) into (\ref{Feq}), the
following equation is obtained for the function $z_{\sigma }(w)$:%
\begin{equation}
\frac{\left[ p(w)z_{\sigma }^{\prime }(w)\right] ^{\prime }}{p(w)}+\left[
2\xi \frac{p^{\prime \prime }(w)}{p(w)}-\frac{u^{2}}{p^{2}(w)}+\lambda ^{2}%
\right] z_{\sigma }(w)=0,  \label{Feq3}
\end{equation}%
where $\lambda ^{2}=\omega ^{2}-m^{2}$, $u=k_{n}$. We denote by $k_{w}$ the
quantum number related to the $w$-direction. The parameter $\lambda $ and
the energy $\omega =\sqrt{\lambda ^{2}+m^{2}}$ depend on that quantum
number. For some geometries it will be convenient to take $\lambda $ in the
role of the quantum number $k_{w}$. The mode functions are specified by the
set of quantum numbers $\sigma =(k_{w},k_{n})$ and $z_{\sigma
}(w)=z_{k_{w},k_{n}}(w)$. The normalization condition for the modes has the
form%
\begin{equation}
\int dw\int_{0}^{2\pi }d\phi \sqrt{|g|}\varphi _{\sigma }^{(\pm )}(x)\varphi
_{\sigma ^{\prime }}^{(\pm )\ast }(x)=\frac{\delta _{nn^{\prime }}}{2\omega }%
\delta _{k_{w}k_{w}^{\prime }}.  \label{norm}
\end{equation}%
Here, $\delta _{k_{w}k_{w}^{\prime }}$ is understood as the Kronecker delta
for discrete quantum number $k_{w}$ and as the Dirac delta function $\delta
(k_{w}-k_{w}^{\prime })$ for continuous one. Substituting (\ref{phisol}) in (%
\ref{norm}), we get the orthonormalization condition for the function $%
z_{k_{w},k_{n}}(w)$:
\begin{equation}
\int dw\,p(w)z_{k_{w},k_{n}}(w)z_{k_{w}^{\prime },k_{n}}^{\ast }(w)=\frac{%
\delta _{k_{w}k_{w}^{\prime }}}{4\pi \omega },  \label{norm2}
\end{equation}%
where the limits of the integration are determined by the range of variation
of the coordinates $w$ and depend on the specific problem (see examples
below).

Having the complete set of mode functions we can evaluate the Hadamard
function, defined as the vacuum expectation value (VEV)%
\begin{equation}
G(x,x^{\prime })=\langle 0|\varphi (x)\varphi ^{\dagger }(x^{\prime
})+\varphi ^{\dagger }(x^{\prime })\varphi (x)|0\rangle .  \label{G1a}
\end{equation}%
Substituting the expansion (\ref{phiexp}) of the field operator and by using
the relations $a_{\sigma },b_{\sigma }|0\rangle =0$, $\langle 0|a_{\sigma
}^{\dagger },b\dagger _{\sigma }=0$, the following mode sum formula is
obtained:%
\begin{equation}
G(x,x^{\prime })=\sum_{\sigma }\sum_{s=+,-}\,\varphi _{\sigma
}^{(s)}(x)\varphi _{\sigma }^{(s)\ast }(x^{\prime }).  \label{G1}
\end{equation}%
Substituting the modes (\ref{phisol}) we get%
\begin{equation}
G(x,x^{\prime })=2\sum_{n=-\infty }^{+\infty }e^{ik_{n}\Delta \phi
}\sum_{k_{w}}\,z_{k_{w},k_{n}}(w)z_{k_{w},k_{n}}^{\ast }(w^{\prime })\cos
(\omega \Delta t),  \label{G12}
\end{equation}%
with $\Delta \phi =\phi -\phi ^{\prime }$, $\Delta t=t-t^{\prime }$, $\omega
=\sqrt{\lambda ^{2}+m^{2}}$, and $\lambda =\lambda (k_{w})$. The symbol $%
\sum_{k_{w}}$ is understood as summation for discrete quantum number $k_{w}$
and integration over $k_{w}$ for the continuous one.

In order to see the effects of the compactification, it is of interest to
compare the function $G(x,x^{\prime })$ with the corresponding function in
the geometry where the coordinate $\phi $ is not compactified, $-\infty
<\phi <+\infty $. The line element is still given by (\ref{ds33}). Now the
mode functions have the form%
\begin{equation}
\varphi _{0\sigma }^{(\pm )}=e^{iu\phi \mp i\omega
t}z_{k_{w},u}^{(0)}(w),\;-\infty <u<+\infty ,  \label{phi0sol}
\end{equation}%
with $\sigma =(k_{w},u)$. The function $z_{k_{w},u}^{(0)}(w)$ obeys the
equation (\ref{Feq3}). From the corresponding normalization condition one
gets%
\begin{equation}
\int dw\,p(w)z_{k_{w},u}^{(0)}(w)z_{k_{w}^{\prime },u}^{(0)\ast }(w)=\frac{%
\delta _{k_{w}k_{w}^{\prime }}}{4\pi \omega },  \label{nc01}
\end{equation}%
which is the same as in (\ref{norm2}). The corresponding Hadamard function
is presented in the form%
\begin{equation}
G_{0}(x,x^{\prime })=2\int_{-\infty }^{+\infty }du\,e^{iu\Delta \phi
}\sum_{k_{w}}\,z_{k_{w},u}^{(0)}(w)z_{k_{w},u}^{(0)\ast }(w^{\prime })\cos
(\omega \Delta t),  \label{G01}
\end{equation}%
again with $\omega =\sqrt{\lambda ^{2}+m^{2}}$.

In the equation (\ref{Feq3}) for $z_{\sigma }(w)$ the parameter $u=k_{n}$
appears in the form $k_{n}^{2}$ and it can be assumed that $z_{\sigma
}(w)=z_{k_{w},k_{n}}(w)=z_{k_{w}}(w,|k_{n}|)$. We apply to the sum over $n$
in (\ref{G12}) the summation formula \cite{Bell10}
\begin{equation}
\sum_{n=-\infty }^{+\infty }g(k_{n})f(|k_{n}|)=\int_{0}^{\infty
}du[g(u)+g(-u)]f(u)+i\int_{0}^{\infty }du[f(iu)-f(-iu)]\sum_{s=\pm 1}\frac{%
g(isu)}{e^{2\pi u+is\alpha _{p}}-1}.  \label{SumForm}
\end{equation}%
with $g(u)=e^{iu\Delta \phi }$ and%
\begin{equation}
f(u)=z_{k_{w}}(w,u)z_{k_{w}}^{\ast }(w^{\prime },u).  \label{fu}
\end{equation}%
It is worth mentioning that the complex conjugate in (\ref{fu}) is taken for
real values of $u$ and then the result is analytically continued in the
complex plane. We will assume that the function obtained in this way is
analytic in the right half-plane. The part with the first term in the
right-hand side gives the Hadamard function (\ref{G01}) and we get%
\begin{eqnarray}
G(x,x^{\prime }) &=&G_{0}(x,x^{\prime })+2i\sum_{k_{w}}\cos (\omega \Delta
t)\,\int_{0}^{\infty }du\sum_{s=\pm 1}\frac{e^{-su\Delta \phi }}{e^{2\pi
u+is\alpha _{p}}-1}  \notag \\
&&\times \lbrack z_{k_{w}}(w,iu)z_{k_{w}}^{\ast }(w^{\prime
},iu)-z_{k_{w}}(w,-iu)z_{k_{w}}^{\ast }(w^{\prime },-iu)].  \label{G1dec}
\end{eqnarray}%
Note that%
\begin{equation}
\sum_{s=\pm 1}\frac{e^{-su\Delta \phi }}{e^{2\pi u+is\alpha _{p}}-1}=\frac{%
2e^{2\pi u}\cosh (2\pi \Delta \phi +i\alpha _{p})-2\cosh (u\Delta \phi )}{%
e^{4\pi u}-2e^{2\pi u}\cos \alpha _{p}+1}.  \label{rel}
\end{equation}
Again, it should be noted that in (\ref{G1dec}) the functions $%
z_{k_{w}}^{\ast }(w^{\prime },\pm iu)$ are understood as the following: we
first consider the function $z_{k_{w}}^{\ast }(w^{\prime },u)$ for real $u$
and then replace $u$ by $\pm iu$. Examples of the function $p(w)$ for which
the solution $z_{k_{w}}(w,u)$ is expressed in terms of the known functions
are given in Section \ref{sec:Special}.

The function $z_{\sigma }(w)=z_{k_{w}}(w,u)$ in (\ref{G1dec}) obeys the
equation (\ref{Feq3}). From here it follows that for any two solutions $%
z_{(1)k_{w}}(w,u)$ and $z_{(2)k_{w}}(w,u)$ of (\ref{Feq3}) the following
Wronskian relation takes place%
\begin{equation}
z_{(1)k_{w}}(w,u)\partial _{w}z_{(2)k_{w}}(w,u)-z_{(2)k_{w}}(w,u)\partial
_{w}z_{(1)k_{w}}(w,u)=\frac{C}{p(w)},  \label{Wronsk}
\end{equation}%
with a constant $C$ which may depend on $k_{w}$ and $u$. This relation is
valid for general $u$, including the complex values.

The representation (\ref{G1dec}) separates explicitly the contribution in
the Hadamard function induced by the compactification of the coordinate $%
\phi $. The local geometry for the compactified and decompactified spaces is
the same and, hence, the corresponding divergences in the coincidence limit $%
x^{\prime }\rightarrow x$ are the same as well. From here it follows that in
the evaluation of the VEVs for local physical observables, biliniear in the
field operator, the renormalization is required for the parts coming from
the function $G_{0}(x,x^{\prime })$. The renormalized topological
contributions in the VEVs are directly obtained from the second term in the
right-hand side of (\ref{G1dec}) or from its derivatives taking the
coincidence limit. In the next section this will be illustrated for the
current density.

\section{VEV of the current density}

\label{sec:Current}

Given the Hadamard function, the VEVs of physical observables bilinear in
the field can be evaluated. As an important characteristic of the vacuum
state here we consider the VEV of the current density $\langle
0|j_{k}(x)|0\rangle \equiv \langle j_{k}(x)\rangle $, where the
corresponding operator is given by
\begin{equation}
j_{k}(x)=ie[\varphi ^{\dagger }(x)\partial _{k}\varphi (x)-(\partial
_{k}\varphi (x))^{\dagger }\varphi (x)].  \label{jl}
\end{equation}%
The VEV is expressed in terms of the Hadamard function as%
\begin{equation}
\langle j_{k}(x)\rangle =\frac{i}{2}e\lim_{x^{\prime }\rightarrow
x}(\partial _{k}-\partial _{k}^{\prime })G(x,x^{\prime }).  \label{jl1}
\end{equation}
With the Hadamard function from (\ref{G1dec}), we see that the charge
density vanishes, $\langle j_{0}\rangle =0$.

For the contravariant component of the current density along the $w$%
-direction, $\langle j^{1}\rangle =-\langle j_{1}\rangle $, we get%
\begin{equation}
\langle j^{1}\rangle =\langle j^{1}\rangle _{0}+\frac{ie}{p(w)}%
\sum_{k_{w}}\,\int_{0}^{\infty }du\sum_{s=\pm 1}\frac{%
C_{k_{w}}(iu)-C_{k_{w}}(-iu)}{e^{2\pi u+is\alpha _{p}}-1},  \label{j1}
\end{equation}%
where%
\begin{equation}
\langle j^{1}\rangle _{0}=\frac{e}{p(w)}\sum_{k_{w}}\int_{-\infty }^{+\infty
}du\,C_{k_{w}}(u)  \label{j10}
\end{equation}%
is the current density in the geometry where the coordinate $\phi $ is not
compactified. Here, the function $C_{k_{w}}(u)$ for real $u$ is defined in
accordance with
\begin{equation}
z_{k_{w}}^{\ast }(w,u)\partial _{w}z_{k_{w}}(w,u)-z_{k_{w}}(w,u)\partial
_{w}z_{k_{w}}^{\ast }(w,u)=\frac{iC_{k_{w}}(u)}{p(w)}.  \label{Clam}
\end{equation}%
For real $u$ the function $C_{k_{w}}(u)$ is real. As it has been emphasized
above, the renormalization is required only for the part $\langle
j^{1}\rangle _{0}$.

By using (\ref{G1dec}) and (\ref{jl1}), for the physical component of the
azimuthal current density, $\langle j^{\phi }\rangle =p(w)\langle
j^{2}\rangle =-\langle j_{2}\rangle /p(w)$, one finds
\begin{equation}
\langle j^{\phi }\rangle =-\frac{2e}{p(w)}\sum_{k_{w}}\int_{0}^{\infty
}du\,u\sum_{s=\pm 1}s\frac{z_{k_{w}}(w,iu)z_{k_{w}}^{\ast
}(w,iu)-z_{k_{w}}(w,-iu)z_{k_{w}}^{\ast }(w,-iu)}{e^{2\pi u+is\alpha _{p}}-1}%
.  \label{jphi}
\end{equation}%
The azimuthal current vanishes in the geometry with uncompactified $\phi $
and the current (\ref{jphi}) is induced by the compatification. The
component $\langle j^{1}\rangle $ is an even function of $\alpha _{p}$ and
the component $\langle j^{\phi }\rangle $ is an odd function. An alternative
expression is obtained from (\ref{jphi}) by using the expansion $%
1/(e^{y}-1)=\sum_{l=1}^{\infty }e^{-ly}$:%
\begin{equation}
\langle j^{\phi }\rangle =\frac{4ie}{p(w)}\sum_{l=1}^{\infty }\sin \left(
l\alpha _{p}\right) \sum_{k_{w}}\int_{0}^{\infty }du\,ue^{-2\pi lu}\left[
z_{k_{w}}(w,iu)z_{k_{w}}^{\ast }(w,iu)-z_{k_{w}}(w,-iu)z_{k_{w}}^{\ast
}(w,-iu)\right] .  \label{jphi2}
\end{equation}%
This current is an odd periodic function of the magnetic flux, enclosed by
the tube, with the period of flux quantum.

Note that, unlike to the vacuum and particle concepts, the expectation value
(\ref{jl1}) is a vector quantity and can be transformed to other reference
frames, with the coordinates $\bar{x}^{i}$, by the standard relation $%
\langle \bar{j}_{i}(\bar{x})\rangle =(\partial x^{k}/\partial \bar{x}%
^{i})\langle j_{k}(x)\rangle $. In general, the vacuum state $|\bar{0}%
\rangle $ realized by the modes $\bar{\varphi}_{\chi }^{(\pm )}(\bar{x})$ in
the new reference frame differ from the vacuum $|0\rangle $. In this case
the mean current density $\langle \bar{j}_{i}(\bar{x})\rangle $ does not
coincide with the vacuum expectation value, $\langle \bar{j}_{i}(\bar{x}%
)\rangle \neq \langle \bar{0}|\bar{j}_{i}(\bar{x})|\bar{0}\rangle $. The
quantity $\langle \bar{j}_{i}(\bar{x})\rangle $ presents the expectation
value of the current density in a state that contains particles of the modes
$\bar{\varphi}_{\chi }^{(\pm )}(\bar{x})$. The mean number of those
particles is determined by the Bogolubov coefficient $\beta _{\chi \sigma
}^{(s)}$.

\section{Vacuum currents in special cases}

\label{sec:Special}

As applications of the general results given above, in this section we
consider special cases.

\subsection{Cylinder with a constant radius}

For a cylinder with a constant radius $L$ the mode functions have the form%
\begin{equation}
z_{\sigma }(w)=z_{k_{w}}(w,|k_{n}|)=\frac{e^{ik_{w}w}}{\sqrt{8\pi
^{2}L\omega }},\;-\infty <k_{w}<+\infty ,  \label{zcyl}
\end{equation}%
with $\omega =\sqrt{k_{w}^{2}+k_{n}^{2}/L^{2}+m^{2}}$. Substituting in (\ref%
{G12}) with $\sum_{k_{w}}\rightarrow \int_{-\infty }^{+\infty }dk_{w}$ and
integrating over $k_{w}$, we get%
\begin{equation}
G(x,x^{\prime })=\frac{1}{2\pi ^{2}L}\sum_{n=-\infty }^{+\infty
}e^{ik_{n}\Delta \phi }K_{0}\left( \sqrt{k_{n}^{2}/L^{2}+m^{2}}\sqrt{\left(
\Delta w\right) ^{2}-\left( \Delta t\right) ^{2}}\right) ,  \label{Gcyl}
\end{equation}%
where $K_{\nu }(x)$ is the Macdonald function \cite{Abra72}. For the
function $C_{k_{w}}(u) $ defined by (\ref{Clam}) one has $%
C_{k_{w}}(u)=k_{w}/(4\pi ^{2}\omega )$. It is an odd function of $k_{w}$ and
the integration over $k_{w}$ gives a zero current along the $w$-direction, $%
\langle j^{1}\rangle =0 $. For the azimuthal current density from (\ref%
{jphi2}) we find
\begin{equation}
\langle j^{\phi }\rangle =\frac{e}{4\pi ^{3}L^{2}}\sum_{l=1}^{\infty }\frac{%
1+2\pi lLm}{l^{2}e^{2\pi lLm}}\sin \left( l\alpha _{p}\right) .
\label{jphicyl}
\end{equation}%
This result is the special case of the general formula from \cite{Beze13T}.

\subsection{Current density on a cone}

In the case of a cone with an opening angle $2\pi \alpha $ the function $%
p(w) $ is given by (\ref{pc}). As a quantum number $k_{w}$ we take $%
k_{w}=\lambda $. The general solution of the equation (\ref{Feq3}) is a
linear combination of the Bessel and Neumann functions $J_{|k_{n}|/\alpha
}(\lambda w)$ and $Y_{|k_{n}|/\alpha }(\lambda w)$. For $|n|\geq 2$, from
the normalizability condition it follows that the function $%
J_{|k_{n}|/\alpha }(\lambda w)$ should be taken. For the modes with $|n|=1$,
under the condition $1-|\alpha _{p}|/2\pi <\alpha $, and for $n=0$, under
the condition $|\alpha _{p}|<2\pi \alpha $, the irregular part with the
function $Y_{|k_{n}|/\alpha }(\lambda w)$ may also be present. In those
cases, in order to uniquely define the mode functions, an additional
boundary condition is required on the cone apex. Here we will consider a
special case where the function
\begin{equation}
z_{k_{w}}(w,|k_{n}|)=c_{1}J_{|k_{n}|/\alpha }(\lambda w),\;0\leq
k_{w}=\lambda <\infty  \label{zcone1}
\end{equation}%
is taken for all values of $n$. By using the integral%
\begin{equation}
\int_{0}^{\infty }dw\,wJ_{|k_{n}|/\alpha }(\lambda w)J_{|k_{n}|/\alpha
}(\lambda ^{\prime }w)=\frac{1}{\lambda }\delta (\lambda -\lambda ^{\prime
}),  \label{Jint}
\end{equation}%
from (\ref{norm2}) for the normalization coefficient one gets $c_{1}=\sqrt{%
\lambda /(4\pi \alpha \omega )}$. With this coefficient, the function $%
z_{k_{w}}(w,u)$ in (\ref{Clam})\ is real and, hence, $C_{k_{w}}(u)=0$. This
shows that $\langle j^{1}\rangle =\langle j^{1}\rangle _{0}=0$ for the
current density in the axial direction.

Substituting the functions (\ref{zcone1}) in (\ref{jphi}), for the VEV of
the azimuthal current we find%
\begin{equation}
\langle j^{\phi }\rangle =\frac{2e}{\pi w}\int_{0}^{\infty }d\lambda \frac{%
\lambda }{\sqrt{\lambda ^{2}+m^{2}}}\int_{0}^{\infty }du\,u\,\mathrm{Im}%
\left( \frac{1}{e^{2\pi \alpha u+i\alpha _{p}}-1}\right) \mathrm{Im}\left[
J_{iu}^{2}(\lambda w)\right] ,  \label{jc1}
\end{equation}%
where the integration variable is redefined as $u/\alpha \rightarrow u$. For
the further transformation we employ the integral representation%
\begin{equation}
\frac{1}{\sqrt{\lambda ^{2}+m^{2}}}=\frac{2}{\sqrt{\pi }}\int_{0}^{\infty
}d\tau e^{-\tau ^{2}(\lambda ^{2}+m^{2})}\ .  \label{first-id}
\end{equation}%
With this representation, the integral over $\lambda $ in (\ref{jc1}) is
evaluated by using the formula from \cite{Prud2}. This gives%
\begin{equation}
\int_{0}^{\infty }d\lambda \lambda e^{-\tau ^{2}\lambda ^{2}}\mathrm{Im}%
\left[ J_{iu}^{2}(\lambda w)\right] =-\frac{\sinh \left( \pi u\right) }{2\pi
\tau ^{2}}e^{-x}K_{iu}(x),  \label{IntBes}
\end{equation}%
with $x=w^{2}/(2\tau ^{2})$. Passing from $\tau $ to a new integration
variable $x$ one gets%
\begin{equation}
\langle j^{\phi }\rangle =\frac{e\sin \alpha _{p}}{\sqrt{2}\pi ^{5/2}w^{2}}%
\int_{0}^{\infty }\frac{dx}{\sqrt{x}}e^{-x-w^{2}m^{2}/2x}\int_{0}^{\infty
}du\,\frac{u\sinh \left( \pi u\right) K_{iu}(x)}{\cosh (2\alpha u)-\cos
\alpha _{p}},  \label{jc2}
\end{equation}%
where we have used
\begin{equation}
\mathrm{Im}\left( \frac{1}{e^{2\alpha u+i\alpha _{p}}-1}\right) =-\frac{1}{2}%
\frac{\sin \alpha _{p}}{\cosh (2\alpha u)-\cos \alpha _{p}}.  \label{Im}
\end{equation}

An alternative expression for the current density in general number of
spatial dimensions $D$ is given in \cite{Beze15}. Specified for the case $%
D=2 $ it has the form%
\begin{eqnarray}
\langle j^{\phi }\rangle &=&\frac{ewm^{3}}{\pi }\left[ \sum_{n=1}^{[1/2%
\alpha ]}\sin (n\alpha _{p})\sin (2\pi n\alpha )g(2mw\sin (\pi n\alpha
))\right.  \notag \\
&&\left. +\frac{1}{2\pi \alpha }\int_{0}^{\infty }dy\,f(\alpha ,\alpha
_{p},y)\frac{g(2mw\cosh (y/2))\sinh y}{\cosh (y/\alpha )-\cos (\pi /\alpha )}%
\right] ,  \label{jcalt}
\end{eqnarray}%
where $[1/2\alpha ]$ stands for the integer part of $1/2\alpha $, $%
g(x)=(1+x)e^{-x}/x^{3}$ and
\begin{equation}
f(\alpha ,\alpha _{p},y)=\sin (\alpha _{p}/2\alpha )\sinh [(1-\alpha
_{p}/2\pi )y/\alpha ]-\sin [\pi (1-\alpha _{p}/2\pi )/\alpha ]\sinh (\alpha
_{p}y/2\pi \alpha ).  \label{galf}
\end{equation}%
For $\alpha >1/2$ the first term in the square brackets of (\ref{jcalt}) is
absent.

The formula (\ref{jc2}) is further simplified for a massless field by using
the integration formula \cite{Prud2}%
\begin{equation}
\int_{0}^{\infty }\frac{dx}{\sqrt{x}}e^{-x}K_{iu}(x)=\sqrt{\frac{\pi }{2}}%
\frac{\pi }{\cosh \left( \pi u\right) }.  \label{IntK}
\end{equation}%
This gives%
\begin{equation}
\langle j^{\phi }\rangle |_{m=0}=\frac{e\sin \alpha _{p}}{2\pi ^{3}w^{2}}%
\int_{0}^{\infty }du\,\frac{u\tanh u}{\cosh (2\alpha u)-\cos \alpha _{p}}.
\label{jcm0}
\end{equation}%
The right-hand side of (\ref{jcm0}) describes the leading behavior of the
current density for a massive field near the cone apex: $\langle j^{\phi
}\rangle \approx \langle j^{\phi }\rangle |_{m=0}$ for $mw\ll 1$. For
massive fields and at large distances, assuming $mw\gg 1$, the current
density is suppressed by the factor $e^{-2mw\sin (\pi \alpha )}$ for $%
1/2\leq \alpha <1$ and by the factor $e^{-2mw}$ in the case $\alpha \leq 1/2$%
.

\section{Hadamard function and the current density on the Beltrami
pseudosphere}

\label{sec:Beltrami}

In this section we specify the general results described above for the
curved geometry corresponding to the Beltrami pseudosphere.

\subsection{Mode functions}

For the Beltrami pseudosphere the line element has the form (\ref{ds2BPS}).
Introducing a new coordinate $r$ in accordance with%
\begin{equation}
r=ae^{-w/a},\;0\leq r<\infty ,  \label{xcc}
\end{equation}%
the corresponding expression takes the form
\begin{equation}
ds^{2}=dt^{2}-\frac{a^{2}}{r^{2}}\left( dr^{2}+L^{2}d\phi ^{2}\right) .
\label{ds3cc2}
\end{equation}%
For the part of the manifold that can be embedded in a 3-dimensional
Euclidean space one has $r\geq L$. Note that the geometry (\ref{ds3cc2}) is
conformally related to the 2D Rinlder spacetime described by the line
element $ds_{\mathrm{R}}^{2}=r^{2}d\tau ^{2}-dr^{2}-dy^{2}$ with
dimensionless time coordinate $\tau =t/a$ and compact coordinate $y=L\phi $:%
\begin{equation}
ds^{2}=\frac{a^{2}}{r^{2}}ds_{\mathrm{R}}^{2}.  \label{RelRind}
\end{equation}%
The current densities in Rinlder spacetime with a toroidally compact
subspace in general number of spatial dimensions have been recently
investigated in \cite{Kota22}.

For the Beltrami pseudosphere the equation (\ref{Feq3}) is rewritten as%
\begin{equation}
z_{\sigma }^{\prime \prime }(r)+\left( \frac{2\xi +a^{2}\lambda ^{2}}{r^{2}}-%
\frac{k_{n}^{2}}{L^{2}}\right) z_{\sigma }(r)=0,  \label{eqcc2}
\end{equation}%
with the general solution%
\begin{equation}
z_{\sigma }(r)=\sqrt{y}\left[ c_{1}I_{i\nu }(y)+c_{2}K_{i\nu }(y)\right]
,\;y=|k_{n}|r/L\,.  \label{zB}
\end{equation}%
Here $I_{\mu }(y)$ is the modified Bessel function and $\nu =\sqrt{%
a^{2}\lambda ^{2}+2\xi -1/4}$. As quantum numbers specifying the mode
functions we can take the set $\sigma =(\nu ,n)$. For the energy of a given
mode we get
\begin{equation}
\omega =\frac{1}{a}\sqrt{\nu ^{2}+\nu _{m}^{2}},  \label{omcc2}
\end{equation}%
where%
\begin{equation}
\nu _{m}^{2}=m^{2}a^{2}+1/4-2\xi .  \label{num}
\end{equation}%
For a conformally coupled field one has $\nu _{m}=ma$.

From the normalizability condition for the mode function it follows that $%
c_{1}=0$. The normalization condition (\ref{norm2}), written in terms of the
coordinate $r$, takes the form%
\begin{equation}
2\pi La^{2}|c_{2}|^{2}\int_{0}^{\infty }\frac{dy}{y}\,K_{i\nu }(y)K_{i\nu
^{\prime }}(y)=\frac{\delta (\nu -\nu ^{\prime })}{2\omega }.  \label{nccc3}
\end{equation}%
By taking unto account that $K_{-i\nu }(y)=K_{i\nu }(y)$, for the quantum
number $\nu $ one has $0<\nu <\infty $. Note that for the values of the
curvature coupling parameter in the range $2\xi >m^{2}a^{2}+1/4$ the energy
becomes imaginary for the modes with $\nu ^{2}<2\xi -m^{2}a^{2}-1/4$. This
leads to the instability of the vacuum state. In the discussion below we
will assume that $2\xi \leq m^{2}a^{2}+1/4$. In particular, this condition
is satisfied by the most important special cases of minimally ($\xi =0$) and
conformally ($\xi =1/8$) coupled fields. By using the integration formula
\cite{Cand76}
\begin{equation}
\int_{0}^{\infty }\frac{dy}{y}\,K_{i\nu }(y)K_{i\nu ^{\prime }}(y)=\frac{\pi
^{2}\delta \left( \nu -\nu ^{\prime }\right) }{2\nu \sinh (\nu \pi )},
\label{KKint}
\end{equation}%
for the normalization coefficient we find%
\begin{equation}
c_{2}=\sqrt{\frac{\nu \sinh (\nu \pi )}{2\pi ^{3}La^{2}\omega }}.
\label{c2cc}
\end{equation}%
Hence, the normalized mode functions (\ref{phisol}) take the form
\begin{equation}
\varphi _{\sigma }^{(\pm )}(x)=\sqrt{\frac{\nu \sinh (\nu \pi )y}{2\pi
^{3}La^{2}\omega }}K_{i\nu }(y)e^{ik_{n}\phi \mp i\omega t},\;y=|k_{n}|r/L,
\label{phiBelt}
\end{equation}%
with $k_{w}=\nu $ and $\omega $ given by (\ref{omcc2}). Note that the
corresponding function
\begin{equation}
z_{\sigma }(w)=z_{k_{w}}(w,|k_{n}|)=c_{2}\sqrt{y}K_{i\nu }(y)  \label{zB2}
\end{equation}%
is real.

\subsection{Hadamard function}

Having the mode functions, for the Hadamard function we get%
\begin{eqnarray}
G(x,x^{\prime }) &=&\frac{\sqrt{rr^{\prime }}}{\pi ^{3}La^{2}}%
\sum_{n=-\infty }^{+\infty }e^{ik_{n}\Delta \phi }\int_{0}^{\infty }d\nu \,%
\frac{\nu }{\omega }\sinh (\nu \pi )  \notag \\
&&\times K_{i\nu }(|k_{n}|r/L)K_{i\nu }(|k_{n}|r^{\prime }/L)\cos (\omega
\Delta t),  \label{G1cc}
\end{eqnarray}%
with $\omega $ from (\ref{omcc2}). For the corresponding function in the
geometry (\ref{ds3cc2}) with uncompactified coordinate $\phi $, $-\infty
<\phi <+\infty $, one has%
\begin{equation}
G_{0}(x,x^{\prime })=\frac{2\sqrt{rr^{\prime }}}{\pi ^{3}La^{2}}%
\int_{0}^{\infty }du\,\cos (u\Delta \phi )\int_{0}^{\infty }d\nu \,\frac{\nu
}{\omega }\sinh (\nu \pi )K_{i\nu }(ur/L)K_{i\nu }(ur^{\prime }/L)\cos
(\omega \Delta t).  \label{G10cc}
\end{equation}%
The integral over $u$ is evaluated by using the formula \cite{Prud2}%
\begin{equation}
\int_{0}^{\infty }du\,\cos (u\Delta \phi )K_{i\nu }(ur/L)K_{i\nu
}(ur^{\prime }/L)=\frac{L\pi ^{2}}{4\sqrt{rr^{\prime }}\cosh (\nu \pi )}%
P_{i\nu -1/2}\left( \frac{r^{2}+r^{\prime 2}+(L\Delta \phi )^{2}}{%
2rr^{\prime }}\right) ,  \label{KKint3}
\end{equation}%
where $P_{i\nu -1/2}(x)$ is the Legendre function. This gives%
\begin{equation}
G_{0}(x,x^{\prime })=\frac{a^{-2}}{2\pi }\int_{0}^{\infty }d\nu \,\frac{\nu
}{\omega }\tanh (\nu \pi )\cos (\omega \Delta t)P_{i\nu -1/2}\left( \frac{%
r^{2}+r^{\prime 2}+(L\Delta \phi )^{2}}{2rr^{\prime }}\right) .
\label{G10cc1}
\end{equation}

The Hadamard function is decomposed as (\ref{G1dec}). For the function
appearing in the topological part one has%
\begin{equation}
z_{k_{w}}(w,iu)z_{k_{w}}^{\ast }(w^{\prime
},iu)-z_{k_{w}}(w,-iu)z_{k_{w}}^{\ast }(w^{\prime },-iu)=i\frac{\sqrt{%
rr^{\prime }}}{2\pi La^{2}}\frac{\nu }{\omega }\mathrm{Im}[J_{i\nu
}(ur/L)J_{i\nu }(ur^{\prime }/L)]\,,  \label{zzBP}
\end{equation}%
where we have used the relation
\begin{equation}
\mathrm{Im}[K_{i\nu }(iy)K_{i\nu }(iy^{\prime })]=\frac{\pi ^{2}}{2}\frac{%
\mathrm{Im}[J_{i\nu }(y)J_{i\nu }(y^{\prime })]}{\sinh (\nu \pi )}.
\label{KKJJ}
\end{equation}%
Hence, the Hadamard function is decomposed as
\begin{eqnarray}
G(x,x^{\prime }) &=&G_{0}(x,x^{\prime })-\frac{\sqrt{rr^{\prime }}}{\pi
La^{2}}\int_{0}^{\infty }d\nu \,\frac{\nu }{\omega }\cos (\omega \Delta t)
\notag \\
&&\times \int_{0}^{\infty }du\,\mathrm{Im}[J_{i\nu }(ur/L)J_{i\nu
}(ur^{\prime }/L)]\sum_{s=\pm 1}\frac{e^{-su\Delta \phi }}{e^{2\pi
u+is\alpha _{p}}-1}.  \label{G1deccc}
\end{eqnarray}%
We can further transform this formula by using the expansion $%
1/(e^{y}-1)=\sum_{l=1}^{\infty }e^{-ly}$. With this expansion, the integral
over $u$ is expressed in terms of the Legendre function $Q_{\mu }(x)$:%
\begin{eqnarray}
G(x,x^{\prime }) &=&G_{0}(x,x^{\prime })-\frac{a^{-2}}{\pi ^{2}}%
\sideset{}{'}{\sum}_{l=-\infty }^{+\infty }e^{-li\alpha
_{p}}\int_{0}^{\infty }d\nu \,\frac{\nu }{\omega }\cos (\omega \Delta t)\,
\notag \\
&&\times \mathrm{Im}\left[ Q_{i\nu -1/2}\left( \frac{r^{2}+r^{\prime
2}+L^{2}\left( 2\pi l+s\Delta \phi \right) ^{2}}{2rr^{\prime }}\right) %
\right] ,  \label{Gdeccc2}
\end{eqnarray}%
where the prime on the summation sign means that the term $l=0$ should be
omitted. Now, by using the relation between the Legendre functions \cite%
{Abra72}, we can see that (note that the function $P_{i\nu -1/2}(z)$ is a
real function for real $\nu $ and $z$)%
\begin{equation}
\mathrm{Im}[Q_{i\nu -1/2}(z)]=-\frac{\pi }{2}\tanh \left( \pi \nu \right)
P_{i\nu -1/2}(z).  \label{ImQ}
\end{equation}%
By taking into account the expression (\ref{G10cc}) for the function $%
G_{0}(x,x^{\prime })$, the final expression is written in the form%
\begin{equation}
G(x,x^{\prime })=\frac{a^{-2}}{2\pi }\sum_{l=-\infty }^{\infty }e^{-il\alpha
_{p}}\int_{0}^{\infty }d\nu \,\frac{\nu }{\omega }\tanh \left( \pi \nu
\right) \cos (\omega \Delta t)\,P_{i\nu -1/2}\left( \frac{r^{2}+r^{\prime
2}+L^{2}\left( \Delta \phi +2\pi l\right) ^{2}}{2rr^{\prime }}\right) .
\label{Gcc}
\end{equation}%
where the $l=0$ term corresponds to the function $G_{0}(x,x^{\prime })$.

The formula (\ref{Gcc}) could also be obtained in another way, by using the
integral representation \cite{Wats95}
\begin{equation*}
K_{i\nu }(br)K_{i\nu }(br^{\prime })=\frac{1}{2}\int_{0}^{\infty }\frac{dy}{y%
}\,\exp \left( -\frac{y}{2}-b^{2}\frac{r^{2}+r^{\prime 2}}{2y}\right)
K_{i\nu }(b^{2}rr^{\prime }/y)
\end{equation*}%
for the product of the Macdonald functions in (\ref{G1cc}). This gives%
\begin{eqnarray}
G(x,x^{\prime }) &=&\frac{\sqrt{rr^{\prime }}}{2\pi ^{3}La^{2}}%
\sum_{n=-\infty }^{+\infty }e^{ik_{n}\Delta \phi }\int_{0}^{\infty }d\nu \,%
\frac{\nu }{\omega }\sinh (\nu \pi )\cos (\omega \Delta t)  \notag \\
&&\times \int_{0}^{\infty }\frac{du}{u}\exp \left( -\frac{k_{n}^{2}}{2uL^{2}}%
-u\frac{r^{2}+r^{\prime 2}}{2}\right) K_{i\nu }(rr^{\prime }u),  \label{Gccn}
\end{eqnarray}%
where we have introduced a new integration variable $u=k_{n}^{2}/(L^{2}y)$.
As the next step we apply the Poisson resummation formula to the series over
$n$. The following relation is obtained:
\begin{equation}
\sum_{n=-\infty }^{+\infty }\exp \left( ik_{n}\Delta \phi -\frac{k_{n}^{2}}{%
2uL^{2}}\right) =L\sqrt{2\pi u}\sum_{l=-\infty }^{\infty }\exp \left[
-il\alpha _{p}-\frac{u}{2}L^{2}\left( \Delta \phi +2\pi l\right) ^{2}\right]
.  \label{PR}
\end{equation}%
Substituting this in (\ref{Gccn}), the integral over $u$ is evaluated with
the help of the formula
\begin{equation*}
\int_{0}^{\infty }\frac{du}{\sqrt{u}}e^{-pu}K_{i\nu }(cu)=\frac{\pi
^{3/2}P_{i\nu -1/2}(p/c)}{\sqrt{2c}\cosh (\pi \nu )},
\end{equation*}%
and we obtain the representation (\ref{Gcc}).

Another representation for the Hadamard function is obtained from (\ref{Gccn}%
) by taking into account the relation (\ref{PR}):%
\begin{eqnarray}
G(x,x^{\prime }) &=&\frac{\sqrt{rr^{\prime }/2}}{\pi ^{5/2}a^{2}}%
\sum_{l=-\infty }^{\infty }e^{-il\alpha _{p}}\int_{0}^{\infty }\frac{du}{%
\sqrt{u}}\exp \left[ -\frac{u}{2}\left( r^{2}+r^{\prime 2}+L^{2}\left(
\Delta \phi +2\pi l\right) ^{2}\right) \right]  \notag \\
&&\times \int_{0}^{\infty }d\nu \,\frac{\nu }{\omega }\sinh (\nu \pi )\cos
(\omega \Delta t)K_{i\nu }(rr^{\prime }u),  \label{Gccn2}
\end{eqnarray}%
By using the formula
\begin{equation}
\sinh \left( \pi \nu \right) K_{i\nu }(z)=\frac{\pi i}{2}\left[ I_{i\nu
}(z)-I_{-i\nu }(z)\right] ,  \label{RelK}
\end{equation}%
for the Macdonald function, this gives
\begin{eqnarray}
G(x,x^{\prime }) &=&\frac{i\sqrt{rr^{\prime }}}{\left( 2\pi \right)
^{3/2}a^{2}}\sum_{l=-\infty }^{\infty }e^{-il\alpha _{p}}\int_{0}^{\infty }%
\frac{du}{\sqrt{u}}\exp \left[ -\frac{u}{2}\left( r^{2}+r^{\prime
2}+L^{2}\left( \Delta \phi +2\pi l\right) ^{2}\right) \right]  \notag \\
&&\times \int_{0}^{\infty }d\nu \,\frac{\nu }{\omega }\sinh (\nu \pi )\cos
(\omega \Delta t)\left[ I_{i\nu }(rr^{\prime }u)-I_{-i\nu }(rr^{\prime }u)%
\right] ,  \label{Gccn3}
\end{eqnarray}%
We rotate the integration contour over $\nu $ by the angle $\pi /2$ for the
term with $I_{-i\nu }(xx^{\prime }u)$ and by the angle $-\pi /2$ for the
term with $I_{i\nu }(xx^{\prime }u)$. Assuming that $\nu _{m}^{2}\geq 0$,
for the integral we get
\begin{equation}
\int_{0}^{\infty }d\nu \,\frac{\nu }{\omega }\cos (\omega \Delta t)\left[
I_{i\nu }(rr^{\prime }u)-I_{-i\nu }(rr^{\prime }u)\right] =\frac{2}{i}%
a\int_{\nu _{m}}^{\infty }d\nu \,\frac{\nu \cosh (\sqrt{\nu ^{2}-\nu _{m}^{2}%
}\Delta t/a)}{\sqrt{\nu ^{2}-\nu _{m}^{2}}}I_{\nu }(rr^{\prime }u).
\label{Int3}
\end{equation}%
By using the integration formula \cite{Prud2}%
\begin{equation}
\int_{0}^{\infty }\frac{du}{\sqrt{u}}e^{-bu/2}I_{\nu }(rr^{\prime }u)=\sqrt{%
\frac{2}{\pi rr^{\prime }}}Q_{\nu -1/2}\left( \frac{b}{2rr^{\prime }}\right)
,  \label{Int4}
\end{equation}%
for the Hadamard function we find the representation%
\begin{equation}
G(x,x^{\prime })=\frac{1}{\pi ^{2}a}\sum_{l=-\infty }^{\infty }e^{il\alpha
_{p}}\int_{\nu _{m}}^{\infty }d\nu \,\frac{\nu \cosh (\sqrt{\nu ^{2}-\nu
_{m}^{2}}\Delta t/a)}{\sqrt{\nu ^{2}-\nu _{m}^{2}}}Q_{\nu -1/2}\left( \frac{%
r^{2}+r^{\prime 2}+L^{2}\left( \Delta \phi -2\pi l\right) ^{2}}{2rr^{\prime }%
}\right) .  \label{Gccn4}
\end{equation}%
Note that the Legendre function in this formula is expressed in terms of the
hypergeometric function as%
\begin{equation}
Q_{\nu -1/2}(z)=\frac{\sqrt{\pi }\Gamma \left( \nu +1/2\right) }{\left(
2z\right) ^{\nu +1/2}\Gamma \left( \nu +2\right) }F\left( \frac{\nu +3/2}{2},%
\frac{\nu +1/2}{2};\nu +1,\frac{1}{z^{2}}\right) .  \label{Qhyp}
\end{equation}

As it is seen from (\ref{ds2BPS}), the geometry of cylindrical tube with
constant radius is obtained from the case of the Beltrami pseudosphere in
the limit $a\rightarrow 0$ for fixed $w$. From (\ref{xcc}) it follows that
the coordinate $r$ is large in that limit and for the argument of the
Legendre function in (\ref{Gccn4}) one has%
\begin{equation}
\frac{r^{2}+r^{\prime 2}+L^{2}\left( \Delta \phi -2\pi l\right) ^{2}}{%
2rr^{\prime }}\approx 1+\frac{\left( \Delta x_{l}\right) ^{2}}{2a^{2}}%
,\;\Delta x_{l}=\sqrt{\left( \Delta w\right) ^{2}+L^{2}\left( \Delta \phi
-2\pi l\right) ^{2}}.  \label{ArgLeg}
\end{equation}%
By taking into account that in the limit under consideration $\nu _{m}$ is
large, we use the asymptotic formula for the Legendre function \cite{Olve10}%
\begin{equation}
Q_{\nu -1/2}(\cosh u)\approx \left( \frac{u}{\sinh u}\right)
^{1/2}K_{0}\left( \nu u\right) ,\;\nu \gg 1.  \label{AsLeg}
\end{equation}%
For the function in (\ref{Gccn4}) one has $u\approx a\Delta x_{l}$.
Substituting (\ref{AsLeg}) in (\ref{Gccn4}) and redefining the integration
variable as $a\nu \rightarrow \nu $, in the leading order we get
\begin{equation}
G(x,x^{\prime })\approx \frac{1}{\pi ^{2}}\sum_{l=-\infty }^{\infty
}e^{il\alpha _{p}}\int_{m}^{\infty }d\nu \,\frac{\nu \cosh (\Delta t\sqrt{%
\nu ^{2}-m^{2}})}{\sqrt{\nu ^{2}-m^{2}}}K_{0}\left( \nu \Delta x_{l}\right) .
\label{Glim1}
\end{equation}%
The integral is evaluated by using the formula from \cite{Prud2} with the
result%
\begin{equation}
G(x,x^{\prime })\approx \frac{1}{2\pi }\sum_{l=-\infty }^{\infty }\frac{%
e^{il\alpha _{p}}e^{-m\sqrt{\left( \Delta x_{l}\right) ^{2}-\left( \Delta
t\right) ^{2}}}}{\sqrt{\left( \Delta x_{l}\right) ^{2}-\left( \Delta
t\right) ^{2}}}.  \label{Glim2}
\end{equation}%
The expression on the right-hand side coincides with the Hadamard function
for a cylindrical tube of constant radius (see, for example, \cite{Beze13T}).

\subsection{Current density}

Similar to the case of conical geometry the function $z_{k_{w}}(w,u)$ is
real for real $u$ (see (\ref{zB2})) and $C_{k_{w}}(u)=0$. Hence, the current
density along the axial direction vanishes, $\langle j^{1}\rangle =0$. The
VEV of the current density in the compact dimension is obtained from the
general formula (\ref{jphi}). By using (\ref{zB2}) we get
\begin{equation}
\langle j^{\phi }\rangle =-\frac{er^{2}}{\pi L^{2}a^{2}}\sin \alpha
_{p}\int_{0}^{\infty }d\nu \,\frac{\nu }{\sqrt{\nu ^{2}+\nu _{m}^{2}}}%
\,\int_{0}^{\infty }du\,\frac{u\,\mathrm{Im}[J_{i\nu }^{2}(ur/L)]}{\cosh
(2\pi u)-\cos \alpha _{p}}.  \label{jphicc}
\end{equation}%
As seen, the current density depends on $r$ and $L$ in the form of the ratio
\begin{equation}
\frac{L}{r}=\frac{L}{a}e^{w/a}.  \label{Lr}
\end{equation}%
For a fixed $w$, the proper length of the compact dimension is given by $%
2\pi Le^{w/a}$. The current density is an odd periodic function of the
parameter $\alpha _{p}$ with the period $2\pi $. This corresponds to
periodicity with respect to the magnetic flux with the period of flux
quantum. An alternative representation is obtained by using the formula (\ref%
{Gcc}) for the Hadamard function:%
\begin{equation}
\langle j^{\phi }\rangle =-\frac{2eL}{ra^{2}}\sum_{l=1}^{\infty }l\sin
(l\alpha _{p})\int_{0}^{\infty }d\nu \,\frac{\nu \tanh \left( \pi \nu
\right) }{\sqrt{\nu ^{2}+\nu _{m}^{2}}}\,P_{i\nu -1/2}^{\prime }\left(
1+2\left( \pi lL/r\right) ^{2}\right) ,  \label{jphicc2}
\end{equation}%
where the prime means the derivative with respect to the argument of the
function.

Another representation for the current density is obtained from (\ref{Gccn4}%
) by using (\ref{jl1}):
\begin{equation}
\langle j^{\phi }\rangle =-\frac{4eL}{\pi ra^{2}}\sum_{l=1}^{\infty }l\sin
\left( l\alpha _{p}\right) \int_{\nu _{m}}^{\infty }d\nu \,\nu \frac{Q_{\nu
-1/2}^{\prime }\left( 1+2\left( \pi lL/r\right) ^{2}\right) }{\sqrt{\nu
^{2}-\nu _{m}^{2}}}.  \label{jphicc3}
\end{equation}%
In the limit $a\rightarrow \infty $, with fixed value of $w$, the parameter $%
\nu $ and the coordinate $r$ are large and we use the asymptotic (\ref{AsLeg}%
) for the Legendre function. It can be seen that in the leading order from (%
\ref{jphicc3}) the result (\ref{jphicyl}) is obtained for the current
density in a cylindrical tube with constant radius.

Let us consider the behavior of the current density in the asymptotic
regions of the ratio $r/L$. For $r/L\ll 1$ the argument of the Legendre
function in (\ref{jphicc3}) is large and we use the formula \cite{Olve10}%
\begin{equation}
Q_{\nu -1/2}\left( x\right) \approx \frac{\sqrt{\pi }\Gamma \left( \nu
+1/2\right) }{\Gamma \left( \nu +1\right) \left( 2x\right) ^{\nu +1/2}},
\label{Ql}
\end{equation}%
for $x\gg 1$. With this asymptotic, the dominant contribution to the
integral over $\nu $ in (\ref{jphicc3}) comes from the region near the lower
limit of integration. In the leading order this gives%
\begin{equation}
\langle j^{\phi }\rangle \approx \frac{2e\Gamma \left( \nu _{m}+3/2\right)
(r/2\pi L)^{2\nu _{m}+2}}{\pi \Gamma \left( \nu _{m}\right) \sqrt{\nu
_{m}\ln (2\pi L/r)}a^{2}}\sum_{l=1}^{\infty }\frac{\sin \left( l\alpha
_{p}\right) }{l^{2\nu _{m}+2}},  \label{jsm}
\end{equation}%
and the current density tends to zero like $\left( r/L\right) ^{2\nu _{m}+2}/%
\sqrt{\ln (L/r)}$. In the opposite limit $r/L\gg 1$ the dominant
contribution to the integral in (\ref{jphicc3}) comes from large $\nu $ and
we employ the asymptotic (\ref{AsLeg}). The integral is evaluated by using
the formula from \cite{Prud2} and the leading order term is expressed as%
\begin{equation}
\langle j^{\phi }\rangle \approx \frac{er^{2}}{4\pi ^{3}L^{2}a^{2}}%
\sum_{l=1}^{\infty }\frac{\sin \left( l\alpha _{p}\right) }{l^{2}},\;r\gg L.
\label{jla}
\end{equation}%
For a given $r$ the radius of the tube is given by $aL/r$ and the limit
under consideration corresponds to the region where the radius of the tube
is much smaller than the curvature radius $a$.

For a conformally coupled massless field we have $\nu _{m}=0$ and the
expression for the current density is transformed to%
\begin{equation}
\langle j^{\phi }\rangle =-\frac{4eL}{\pi ra^{2}}\sum_{l=1}^{\infty }l\sin
\left( l\alpha _{p}\right) \int_{0}^{\infty }d\nu \,Q_{\nu -1/2}^{\prime
}\left( 1+2\left( \pi lL/r\right) ^{2}\right) .  \label{jphiccm0}
\end{equation}%
As it has been mentioned above, the geometry at hand is conformally related
to the 2D Rindler spacetime with a compact dimension (see (\ref{RelRind})).
For a conformally coupled massless field we expect the relation%
\begin{equation}
\langle j^{\phi }\rangle =\frac{r^{2}}{a^{2}}\langle j^{\phi }\rangle _{%
\mathrm{R}},  \label{jphiBelRind}
\end{equation}%
where $\langle j^{\phi }\rangle _{\mathrm{R}}$ is the current density in the
Rindler spacetime. The expression for the latter is obtained from the
general formula in \cite{Kota22} taking $D=2$ for spatial dimension and
making the replacement $L\rightarrow 2\pi L$:%
\begin{equation}
\langle j^{\phi }\rangle _{\mathrm{R}}=\frac{eL}{4r^{3}}\sum_{l=1}^{\infty
}l\sin \left( l\alpha _{p}\right) \left\{ \frac{1}{\left( \pi lL/r\right)
^{3}}-\int_{0}^{\infty }\,dx\frac{\left[ \cosh ^{2}x+(\pi lL/r)^{2}\right]
^{-\frac{3}{2}}}{x^{2}+\pi ^{2}/4}\right\} .  \label{jphiRind}
\end{equation}%
We have checked numerically that the following relation is valid:%
\begin{equation}
\int_{0}^{\infty }d\nu \,Q_{\nu -1/2}\left( 1+2u^{2}\right) =\frac{\pi }{4}%
\left[ \frac{1}{u}-\int_{0}^{\infty }d\nu \frac{\left( \cosh ^{2}\nu
+u^{2}\right) ^{-\frac{1}{2}}}{\nu ^{2}+\pi ^{2}/4}\right] .  \label{Ident1}
\end{equation}%
This relation shows that the connection (\ref{jphiBelRind}) indeed takes
place between the current densities on the Beltrami pseudosphere and in 2D
Rindler spacetime with compact dimension.

\begin{figure}[tbph]
\begin{center}
\begin{tabular}{cc}
\epsfig{figure=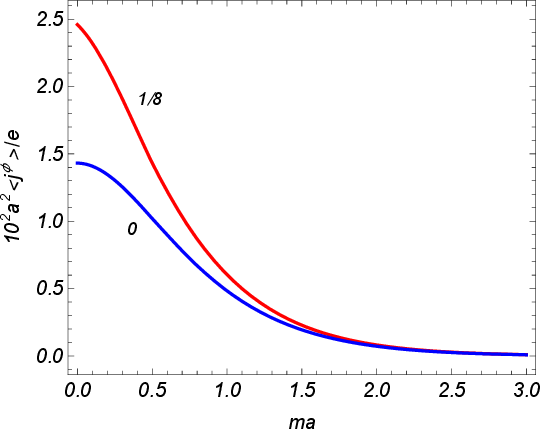,width=7.5cm,height=6cm} & \quad %
\epsfig{figure=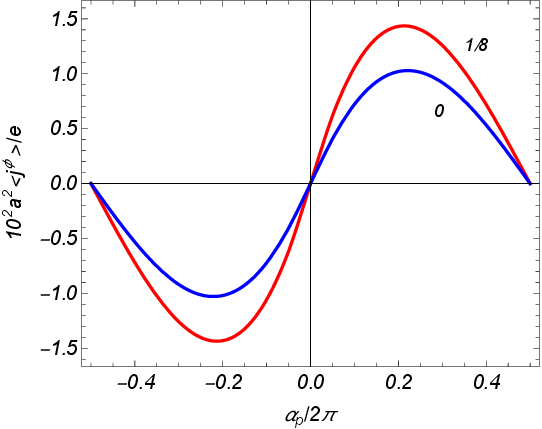,width=7.5cm,height=6cm}%
\end{tabular}%
\end{center}
\caption{The current density as a function of the mass (left panel) and of
the phase in the quasiperiodicity condition (right panel) for $L/r=0.5$. For
the left panel we have taken $\protect\alpha _{p}=2\protect\pi /5$ and for
the right panel $ma=0.5$.}
\label{fig3}
\end{figure}

The left panel of Fig. \ref{fig3} presents the dependence of the current
density on the mass of the field for $L/r=0.5$ and $\alpha _{p}=2\pi /5$.
The right panel displays the current density versus $\alpha _{p}/2\pi $ for
the same value of $L/r$ and for $ma=0.5$. On both panels the numbers near
the curves are the corresponding values of the curvature coupling parameter.

\begin{figure}[tbph]
\begin{center}
\epsfig{figure=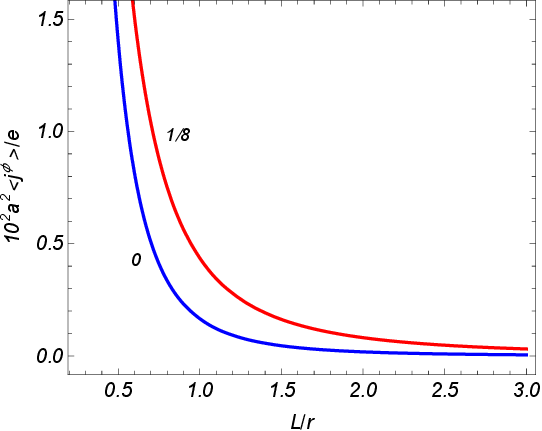,width=8cm,height=7cm}
\end{center}
\caption{The current density versus $L/r$ for conformally ($\protect\xi =1/8$%
) and minimally ($\protect\xi =0$) coupled massless fields and for fixed $%
\protect\alpha _{p}=\protect\pi /2$.}
\label{fig4}
\end{figure}

In Fig. \ref{fig4} we have plotted the dependence of the current density on
the ratio $L/r$ for minimally and conformally coupled massless fields. The
graphs are plotted for $\alpha _{p}=\pi /2$. As was expected, the current
density tends to zero in the limit $L/r\rightarrow \infty $.

\begin{figure}[tbph]
\begin{center}
\epsfig{figure=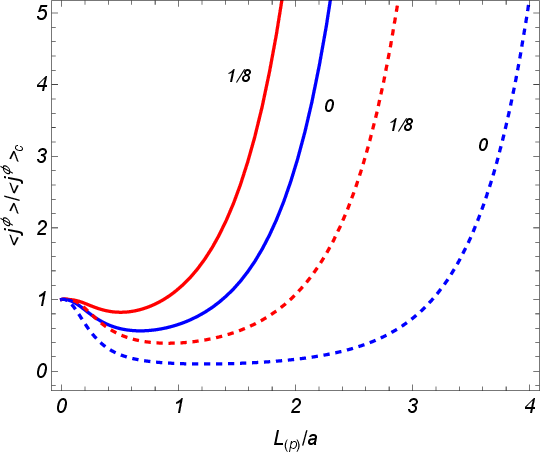,width=8cm,height=7cm}
\end{center}
\caption{The ratio of the current densities on the Beltrami pseudosphere and
on the tube with constant radius (full curves) versus the proper radius of
the tube in units of the curvature radius $a$. The dashed curves present the
same ratio for the AdS tube and the tube of constant radius. The graphs are
plotted for conformally and minimally coupled fields and for $ma=0.5$, $%
\protect\alpha _{p}=\protect\pi /2$.}
\label{fig5}
\end{figure}

It is of interest to compare the vacuum current densities for different
geometries of the 2D tube. The solid curves in Fig. \ref{fig5} present the
ratio of the current densities on the Beltrami pseudosphere and on the tube
with a constant radius, $\langle j^{\phi }\rangle /\langle j^{\phi }\rangle
_{c}$, as a function of the proper radius $L_{(p)}$ of the tube in units of
the curvature radius $a$. For the Beltrami pseudosphere $L_{(p)}=aL/r$ and
for the tube with constant radius $L_{(p)}=L$. The current density for the
tube with a constant radius is given by (\ref{jphicyl}) and the ratio $%
\langle j^{\phi }\rangle /\langle j^{\phi }\rangle _{c}$ is evaluated for
the same values of the proper radii for the Beltrami pseudosphere and
constant radius tube. The dashed lines in Fig. \ref{fig5} present the same
ratio for the AdS tube instead of the Beltrami pseudosphere. The proper
radius for the AdS tube one has $L_{(p)}=aL/z$ and the corresponding current
density is given by (\ref{jphiads}). The graphs in Fig. \ref{fig5} are
plotted for $ma=0.5$, $\alpha _{p}=\pi /2$, and the numbers near the curves
are the corresponding values of the curvature coupling parameter $\xi $. As
seen from the graphs, for large values of the proper radius of the tube one
has $\langle j^{\phi }\rangle /\langle j^{\phi }\rangle _{c}\gg 1$ for both
the Beltrami pseudoshpere and AdS tube. This is related to the fact that in
that range the decay of the current densities for the Beltrami pseudoshpere
and AdS tube, as functions of the proper radius, follows a power-law,
whereas for a constant radius tube the decay is exponential for a massive
field.

\section{Conclusion}

\label{sec:Conc}

We have investigated combined effects of spatial curvature and topology on
the local properties of the ground state for a quantum scalar field in
(2+1)-dimensional spacetime. The corresponding line element has the form (%
\ref{ds33}) and it describes rotationally symmetric 2D curved tube with
general dependence of the radius on the axial coordinate. As special cases,
a cylindrical tube with a constant radius, conical tube and tubes with
constant positive and negative curvature are considered. The local
properties of the vacuum state are encoded in two-point functions. As such,
the Hadamard function is taken. It is evaluated by using the mode-sum
formula (\ref{G1}) with scalar field modes given by (\ref{phisol}) and
obeying the quasiperiodicity condition (\ref{Per}) with a general phase $%
\alpha _{p}$. The dependence of the mode functions on the axial coordinate
is expressed in terms of the function $z_{\sigma }(w)$ being the solution of
the equation (\ref{Feq3}) and normalized by the condition (\ref{norm2}).

In order to extract the effects induced by the compactification, we have
separated from the Hadamard function the part that corresponds to the same
local geometry with decompactified $\phi $-coordinate. The separation is
done by using the Abel-Plana-type summation formula and the decomposed
representation of the Hadamard function is given by (\ref{G1dec}). The VEVs
of local physical observables, bilinear in the field operator, are obtained
from the Hadamard function and its derivatives in the coincidence limit of
the spacetime arguments. That limit is divergent and a renormalization
procedure is required to extract finite physical values of the VEVs. The
advantage of the presentation (\ref{G1dec}) is that the difference $%
G(x,x^{\prime })-G_{0}(x,x^{\prime })$ is finite in the coincidence limit $%
x^{\prime }\rightarrow x$ and the last term can be directly used for the
evaluation of renormalized topological contributions in the VEVs.

As an important physical characteristic of the vacuum state we have
considered the expectation value of the current density. The general
expressions for the currents in the axial and angular directions are given
by (\ref{j1}) and (\ref{jphi}). In the special cases of constant radius and
conical tubes the axial current vanishes and the general formulas for the
current along compact dimension are reduced to the results previously
investigated in the literature. For a conical tube, a new representation (%
\ref{jc2}) is provided. As another application of general formulas, in
Section \ref{sec:Beltrami} the Hadamard function and the current density on
the Beltrami pseudosphere are studied. The corresponding mode functions are
given by (\ref{zB2}) and the Hadamard function is expressed as (\ref{G1cc}).
An alternative expression is provided by formula (\ref{Gccn4}). The
component of the vacuum current along the axial direction vanishes and for
the VEV of the component along the compact dimension we have provided
several representations (see (\ref{jphicc}), (\ref{jphicc2}) and (\ref%
{jphicc3})). The spacetime geometry for the Beltrami pseudosphere is
conformally related to the one for 2D Rindler spacetime (see (\ref{RelRind}%
)) and we have shown that for a conformally coupled massless field the
standard relation between the corresponding VEVs in conformally connected
problems take place. For a massive field with general curvature coupling the
dimensionless combination $a^{2}\langle j^{\phi }\rangle $ is a function of
the ratio $L/r$ which gives the proper radius of compact dimension measured
in units of the curvature radius $a$. For the radius of the tube much
smaller than the curvature radius, $L/r\ll 1$, the effect of the spatial
curvature on the current density is weak and, in the leading order, the VEV
coincides with the corresponding current density for a tube with a constant
radius equal to the proper radius $aL/r$ (see (\ref{jla})). In the opposite
limit of large proper radius, $L/r\gg 1$, the effect of curvature is
essential and the asymptotic is described by (\ref{jsm}). In this limit, the
decay of the current density as a function of the proper radius follows a
power-law with an additional logarithmic factor, like $\left( r/L\right)
^{2\nu _{m}+2}/\sqrt{\ln (L/r)}$. This behavior is in clear contrast to the
one for a constant radius tube when the current density for a massive field
is suppressed exponentially.

\section*{Acknowledgments}

The work was supported by the grant No. 21AG-1C047 of the Higher Education
and Science Committee of the Ministry of Education, Science, Culture and
Sport RA and by the ANSEF grant 23AN:PS-hepth-2889.

\appendix

\section{Vacuum currents in locally dS and AdS tubes}

\label{sec:dStubes}

The vacuum currents for a scalar field in locally dS and AdS spacetimes with
toroidal subspaces for general number of spatial dimension have been
considered in Refs. \cite{Bell13dS} and \cite{Beze15AdS}, respectively (for
a review and further discussions see also \cite{Saha24}). Here we specify
the corresponding results for the case of 2D space with a single compact
dimension (dS and AdS tubes).

For dS tubes the line element reads
\begin{equation}
ds_{\mathrm{dS}}^{2}=dt^{2}-e^{-2t/a}\left( dw^{2}+L^{2}d\phi ^{2}\right) .
\label{dsdS}
\end{equation}%
For the Ricci scalar one has $R=6/a^{2}$ and the nonzero components of the
Ricci tensor are given by (no summation over $i$) $R_{i}^{i}=2/a^{2}$, $%
i=0,1,2$. From the general result of \cite{Bell13dS} for the physical
component of the current density along the compact dimension we get%
\begin{equation}
\langle j^{\phi }\rangle _{\mathrm{dS}}=-\frac{2eL}{a^{2}\eta }%
\sum_{l=1}^{\infty }l\sin (l\alpha _{p})\partial _{u}F_{\nu _{-}}^{\mathrm{%
(dS)}}(u),\;u=2\left( \pi l\frac{L}{\eta }\right) ^{2}-1,  \label{jphidS}
\end{equation}%
where
\begin{equation}
\eta =ae^{t/a},\;\nu _{\pm }=\sqrt{1-6\xi \pm m^{2}a^{2}}.  \label{eta}
\end{equation}%
In (\ref{jphidS}) \ we have defined the function%
\begin{eqnarray}
F_{\nu }^{\mathrm{(dS)}}(u) &=&\frac{\sinh \nu x}{\sin \left( \pi \nu
\right) \sinh x},\;x=\mathrm{arccosh}\,u,\;\text{for }u\geq 1,  \notag \\
F_{\nu }^{\mathrm{(dS)}}(u) &=&\frac{\sin \nu x}{\sin \left( \pi \nu \right)
\sin x},\;x=\mathrm{arccos}\,u,\;\text{for }u\leq 1.  \label{FdS}
\end{eqnarray}%
In deriving (\ref{jphidS}) it was assumed that the field is prepared in the
Bunch-Davies vacuum state. The parameter $\nu $ can be either positive or
purely imaginary. Note that $\tau =-\eta $, $-\infty <\tau <0$ is the
conformal time in terms of which the line element is written in conformally
flat form $ds_{\mathrm{dS}}^{2}=(a/\tau )^{2}\left( d\tau
^{2}-dw^{2}-L^{2}d\phi ^{2}\right) $. In the case of a conformally coupled
massless field one has $\nu _{-}=1/2$ and for the function (\ref{FdS}) we
find $F_{\nu }^{\mathrm{(dS)}}(u)=2^{3/2}/\sqrt{u+1}$. For the current
density this gives%
\begin{equation}
\langle j^{\phi }\rangle _{\mathrm{dS}}=\frac{e\eta ^{2}}{4\pi ^{3}L^{2}a^{2}%
}\sum_{l=1}^{\infty }\frac{\sin (l\alpha _{p})}{l^{2}}.  \label{jdScc}
\end{equation}%
As expected, in this special case we have the conformal relation $\langle
j^{\phi }\rangle _{\mathrm{dS}}=(\eta /a)^{2}\langle j^{\phi }\rangle $,
where $\langle j^{\phi }\rangle $ is the current density for a massless
field on a cylindrical tube with constant radius (see (\ref{jphicyl}) for $%
m=0$).

In the case of AdS tubes the line element has the form%
\begin{equation}
ds_{\mathrm{AdS}}^{2}=e^{2w/a}\left( dt^{2}-L^{2}d\phi ^{2}\right) -dw^{2},
\label{dsads}
\end{equation}%
with the curvature scalar and Ricci tensor $R=-6/a^{2}$ and $%
R_{i}^{i}=-2/a^{2}$ (no summation over $i$). The expression for the vacuum
current density is obtained from the general formula in Ref. \cite{Beze15AdS}%
:%
\begin{equation}
\langle j^{\phi }\rangle _{\mathrm{AdS}}=-\frac{2eL}{a^{2}z}%
\sum_{l=1}^{\infty }l\sin (l\alpha _{p})\partial _{u}F_{\nu _{+}}^{\mathrm{%
(AdS)}}(u),\;u=2\left( \pi l\frac{L}{z}\right) ^{2}+1,  \label{jphiads}
\end{equation}%
with $z=ae^{w/a}$, $0<z<\infty $, and%
\begin{equation}
F_{\nu }^{\mathrm{(AdS)}}(u)=\frac{e^{-\nu x}}{\sinh x},\;x=\mathrm{arccosh}%
\,u.  \label{Fads}
\end{equation}%
With the new axial coordinate $z$, the line element is presented in a
conformally flat form $ds_{\mathrm{dS}}^{2}=(a/z)^{2}\left( d\tau
^{2}-dz^{2}-L^{2}d\phi ^{2}\right) $. The circles $z=0$ and $z=\infty $
correspond to the AdS boundary and horizon.

For a conformally coupled massless field $\nu _{+}=1/2$ and
\begin{equation}
F_{\nu }^{\mathrm{(AdS)}}(u)=\frac{1}{\sqrt{2}}\left( \frac{1}{\sqrt{u-1}}-%
\frac{1}{\sqrt{u+1}}\right) .  \label{Fadscc}
\end{equation}%
Putting this in (\ref{jphiads}) one gets $\langle j^{\phi }\rangle _{\mathrm{%
AdS}}=\left( z/a\right) ^{2}\langle j^{\phi }\rangle _{(1)}$, where%
\begin{equation}
\langle j^{\phi }\rangle _{(1)}=\frac{e}{4\pi ^{3}L^{2}}\sum_{l=1}^{\infty }%
\frac{\sin (l\alpha _{p})}{l^{2}}\left\{ 1-\left[ 1+\left( \frac{z}{\pi lL}%
\right) ^{2}\right] ^{-3/2}\right\} .  \label{jphi1}
\end{equation}%
We have a conformal relations between the current densities on AdS tube and
on a cylindrical tube with constant radius having a circular edge at $z=0$.
On the latter the field obeys the Dirichlet boundary condition. The presence
of the edge in the Minkowski counterpart is related to the boundary
condition for the field on the AdS boundary. The edge $z=0$ in the
Minkowskian problem is the conformal image of the AdS boundary. Hence,
formula (\ref{jphi1}) provides the expression of the current density for a
massless scalar field with the boundary condition $\varphi (x)=0$ on the
edge $z=0$ of a cylindrical surface with constant radius. The effects of
edges of AdS tubes on the vacuum current densities for scalar and fermionic
fields are discussed in \cite{Bell15sc2,Bell16sc,Bell18f,Bell20f}.

\end{document}